\documentclass{aa}  

\usepackage[utf8]{inputenc}
\usepackage[varg]{txfonts}
\usepackage{graphicx}
\usepackage{units}
\usepackage{siunitx}[=v2]
\usepackage{textgreek}
\usepackage[super]{nth}
\usepackage{placeins}

\begin{document}

   \title{
   Pyodine: An open, flexible reduction software \\for iodine-calibrated precise radial velocities\thanks{Radial velocity data of $\sigma$~Draconis and HIP~36616 are only available in electronic form at the CDS via anonymous ftp to \protect\url{cdsarc.cds.unistra.fr} (130.79.128.5) or via \protect\url{https://cdsarc.cds.unistra.fr/cgi-bin/qcat?J/A+A/}.}
   }


   \author{Paul Heeren\inst{1,2}
          \and
          Ren\'e Tronsgaard\inst{2}
          \and
          Frank Grundahl\inst{2}
          \and
          Sabine Reffert\inst{1}
          \and
          Andreas Quirrenbach\inst{1}
          \and 
          Pere L. Pall\'e\inst{3,4}
          }

   \institute{
            Landessternwarte, Zentrum f\"ur Astronomie der Universit\"at Heidelberg, K\"onigstuhl 12, 69117 Heidelberg, Germany\\
            \email{pheeren@lsw.uni-heidelberg.de}
         \and
            Stellar Astrophysics Centre, Department of Physics and Astronomy, Aarhus University, Ny Munkegade 120, DK-8000 Aarhus C, Denmark
        \and
            Instituto de Astrof\'isica de Canarias, E-38200 La Laguna, Tenerife, Spain
        \and
            Universidad de La Laguna (ULL), Departamento de Astrof\'isica, E-38206 La Laguna, Tenerife, Spain
             }

   \date{Received 7 July 2022; accepted 24 April 2023}

 
  \abstract
   {Many telescopes use an iodine (I$_2$) absorption cell to measure precise radial velocities (RVs), but their data reduction pipelines are all tailored to their respective instrumental characteristics and not openly accessible.}
   {For existing and future projects dedicated to measuring precise RVs, we have created an open-source, flexible data reduction software to extract RVs from \'echelle spectra via the I$_2$ cell method. The software, called \texttt{pyodine}, is completely written in Python and has been built in a modular structure to allow for easy adaptation to different instruments.}
   {We present the fundamental concepts employed by \texttt{pyodine}, which build on existing I$_2$ reduction codes, and give an overview of the software's structure. We adapted \texttt{pyodine} to two instruments, Hertzsprung SONG located at Teide Observatory (SONG hereafter) and the Hamilton spectrograph at Lick Observatory (Lick hereafter), and demonstrate the code's flexibility and its performance on spectra from these facilities.
   }
   {Both for SONG and Lick data, the \texttt{pyodine} results generally match the RV precision achieved by the dedicated instrument pipelines. Notably, our code reaches a precision of roughly $\unit[0.69]{m\, s^{-1}}$ on a short-term solar time series of SONG spectra, and confirms the planet-induced RV variations of the star HIP~36616 on spectra from SONG and Lick. Using the solar spectra, we also demonstrate the capabilities of our software in extracting velocity time series from single absorption lines. A probable instrumental effect of SONG is still visible in the \texttt{pyodine} RVs, despite being a bit damped as compared to the original results.}
   {With \texttt{pyodine} we prove the feasibility of a highly precise, yet instrument-flexible I$_2$ reduction software, and in the future the code will be part of the dedicated data reduction pipelines for the SONG network and the Waltz telescope project in Heidelberg.}


   \titlerunning{Pyodine: An open, flexible reduction software for I$_2$-calibrated RVs}
   \authorrunning{P. Heeren et al.}
   \maketitle
%

\section{Introduction}
\label{sec:Introduction}

The radial velocity (RV) method is the second-most successful technique for the detection of exoplanets today, accounting for more than 1000 discoveries, and many more confirmations of planet candidates discovered first through other methods\footnote{\url{https://exoplanetarchive.ipac.caltech.edu/}}. This success has been possible thanks to immense progress on instruments and continuing development of advanced data analysis methods.

Stellar RVs are measured by detecting Doppler-induced shifts of the absorption lines of a star in high-resolution spectra, usually obtained with \'echelle spectrographs. Two different methods for the computation of RVs have been developed in the last decades: The first one aims to keep the instrument in a highly stabilized environment, which allows one to directly make use of the wavelength solution produced by a calibration source (either simultaneously or before and after the observations). The small Doppler shifts of the stellar features over time are then measured either by using a cross-correlation technique with a mask \citep[see e.g.,][]{Baranne1996,Pepe2002}, or through forward modeling of the observed spectra with a template \citep[][]{AngladaEscude2012,Zechmeister2018}. Today, instruments such as CARMENES \citep{Quirrenbach2016}, HARPS \citep{Mayor2003}, and ESPRESSO \citep{Pepe2021} reach an internal RV precision below $\sim \unit[1]{m\, s^{-1}}$ using methods based on this principle, and several high-performance, open-source analysis software packages for the extraction of RVs exist \citep[e.g., CERES, SERVAL, and WOBBLE:][]{Brahm2017,Zechmeister2018,Bedell2019}.

For less stable instruments, where the wavelength solution and line-spread function (LSF) of the spectrograph vary considerably over time, the iodine cell method has proven its merit as the second successful technique for measuring precise RVs. In this method, a temperature-stabilized glass cylinder filled with gaseous molecular iodine (I$_2$) is inserted into the light path between the telescope and the spectrograph; the recorded spectra thus show a combination of stellar and I$_2$ absorption features, the latter of which can be used as a precise wavelength reference. The RVs of the star are then found by forward-modeling the observations from stellar and I$_2$ template spectra with a Doppler shift applied to the stellar template.

While the general idea of using a superimposed reference spectrum for precise RV measurements and first applications of this technique date back to \citet{Griffin1973} and \citet{Campbell1979}, \citet{MarcyButler1992} established I$_2$ as the gas of choice for the absorption cell and further developed the technique in the early 1990s to reach a RV precision of $\sim \unit[3]{m\, s^{-1}}$ \citep{Valenti1995,Butler1996}. More recently, \citet{Butler2017} have demonstrated a precision at the $\sim \unit[1]{m\, s^{-1}}$ level on data collected in a 20-year survey with HIRES/Keck using the I$_2$ cell method, and a similar precision is reached on shorter time scales by Hertzsprung SONG on Tenerife, Spain, on some stars \citep{Andersen2016}.

Due to its less strict requirements on instrument stability, the I$_2$ cell method offers a possibility for measuring precise RVs even for smaller telescope projects with constrained financial budgets. We therefore chose this technique for our own Waltz telescope project at Landessternwarte (LSW) Heidelberg \citep{Tala2016}, which aims to continue the RV survey of evolved stars originally performed at Lick Observatory \citep{Frink2001,Reffert2015}. However, in contrast to the RV measurement method on highly stabilized instruments, to our knowledge no open-source analysis software exists for the I$_2$ cell method. Furthermore, I$_2$ cell codes such as the one by \citet{Butler1996} (called \texttt{Butler} code hereafter) or the SONG reduction code, called \texttt{iSONG} \citep{Corsaro2012,Antoci2013,Grundahl2017}, are tailored to specific instruments and thus lack the flexibility to be easily applicable to other projects.

We therefore decided to develop a new software package for the extraction of precise RVs from spectra obtained with the I$_2$ cell method, with the goals of (i) reaching the RV precision achieved by other analysis codes, and (ii) allowing a quick adaptation to different instruments. In the development, we took guidance from the \texttt{iSONG} code, and from a code maintained by Debra Fischer at Yale University. The resulting software, called \texttt{pyodine}, is written in Python~3, built with a modular, object-oriented approach, and publicly available under MIT license\footnote{\url{https://gitlab.com/Heeren/pyodine}}. It has been adapted to and tested on spectra from two different instruments thus far, namely the Hertzsprung SONG node in Tenerife (SONG hereafter), and the Hamilton spectrograph at Lick Observatory (Lick hereafter), and in both cases the precision of the dedicated instrument pipelines (\texttt{Butler} code and \texttt{iSONG}) is generally met.

In this work, we briefly describe the mathematical concepts used in the code (Section~\ref{sec:TheoreticalBackground}) and explain the fundamental structure and the workflow of the software (Section~\ref{sec:CodeImplementation}). Furthermore, in Section~\ref{sec:Results} we present some examples of results for the RVs generated with \texttt{pyodine} from SONG and Lick spectra, showcasing the performance and flexibility of the software. Finally, we summarize our work and give an outlook on future development on the code (Section~\ref{sec:Conclusions}).

\section{Theoretical background}
\label{sec:TheoreticalBackground}

The general mathematical description of the analysis follows algorithms presented in \citet{MarcyButler1992}, \citet{Valenti1995}, and \citet{Butler1996}; for convenience, we summarize all of the necessary equations in the following subsections.

\subsection{Modeling observations}
\label{subsec:ModellingObservations}

The analysis is performed on reduced \'echelle spectra of the target star, with orders already extracted and a rough wavelength scale, typically based on Thorium-Argon (ThAr) calibration spectra, already known. Each observation of the star must have been performed with an I$_2$ cell in the light path, so that the resulting spectrum is a combination of the stellar and I$_2$ absorption lines. As the I$_2$ spectral features are mostly present in the wavelength range $\unit[5000]{\AA} < \lambda < \unit[6300]{\AA}$, only orders covering these parts of the spectrum are used. Each order is then divided into chunks of roughly $\unit[2]{\AA}$ in length, and each chunk is modeled individually, in order to reduce the overall complexity of the model and account for spatial variations in the instrumental line spread function (LSF).

As the observed spectrum contains both stellar and I$_2$ spectral features, high-resolution template spectra for the I$_2$ and the observed star are required. The I$_2$ template spectrum, $T_\mathrm{I2}(\lambda)$, is typically obtained by scanning the I$_2$ cell in a Fourier-Transform Spectrometer (FTS) of very high resolving power \citep[for a detailed discussion and analysis of I$_2$ template spectra, compare][]{Wang2020}. Retrieving the stellar template spectrum $T_*(\lambda)$ usually involves a more complex routine, which is explained in Section~\ref{subsec:DeconvolvedStellarTemplate}.

We denote the observed spectrum over pixels $x$ of a given chunk as $I_\mathrm{obs}(x)$. This is then fitted in a non-linear least squares approach with the model
\begin{equation}\label{equ:IodineObsModel}
    \hat{I}_\mathrm{obs}(x) = k(x) \, \Big[T_\mathrm{I2}\big(\lambda(x)\big) \cdot T_*\big(\lambda(x) \cdot (1 + z)\big) \Big] * L(x) \quad \mathrm{,}
\end{equation}
which incorporates the following sub-models:

$k(x)$, a model of the continuum flux values in the given chunk (Section~\ref{subsubsec:ContinuumWavelengthModels});

$T_\mathrm{I2}\big(\lambda(x)\big)$, the part of the I$_2$ template spectrum corresponding to the modeled chunk;

$\lambda(x)$, a model of the wavelength scale in the chunk (Section~\ref{subsubsec:ContinuumWavelengthModels});

$T_*\big(\lambda(x) \cdot (1+z)\big)$, the part of the stellar template spectrum corresponding to the modeled chunk, taking the (relativistic) Doppler-shift by the relative RV between observation and template $z = [(1 + \beta) / (1 - \beta)]^{1/2}$ into account, with $\beta = v/c$, where $v$ is the RV estimate of that chunk (including the barycentric velocity of the observatory; Section~\ref{subsec:DeconvolvedStellarTemplate});

and $L(x)$, a model of the instrumental line-spread function (LSF; Section~\ref{subsubsec:LSFmodel}).

The philosophy behind the combined model presented in Equation~\ref{equ:IodineObsModel} is that spatial and temporal variations of the instrumental LSF are directly taken into account, given that the chosen LSF description $L(x)$ is flexible enough \citep[for a more thorough discussion compare][]{MarcyButler1992,Valenti1995}. The (relative) RV of an observation is finally computed through a weighted mean of the best-fit velocities $v$ of all chunks (see Section~\ref{subsec:VelocityCombination}).

We note that the combined model is first built on an oversampled pixel grid (usually by a factor $4$ to $6$ with respect to the observation spectrum), to ensure good convolution and to preserve the high-frequency information from the iodine cell; for evaluation by the fitting routine the model is then re-binned to the observation pixels.

\subsubsection{The LSF model}
\label{subsubsec:LSFmodel}

In its simplest form, the instrumental LSF can be described by a model consisting of a single Gaussian function:
\begin{equation}\label{equ:LSFModelSingle}
    L_\mathrm{Single}(x) = \exp \left(-\frac{x^2}{2 c^2}\right) \quad \mathrm{,}
\end{equation}
where the width $c$ is a free model parameter. For most instruments, this model will not deliver a very good description of the actual spectrograph LSF, as asymmetries cannot be accounted for. However, it can be useful in order to determine an estimate of other free parameters (e.g., for the continuum and wavelength models), which can then be used as initial guesses for a more complex modeling.

A more flexible parametrization of a spectrograph LSF has been developed by \citet{Valenti1995}, who use a model built from a central Gaussian and $n/2$ satellite Gaussians on either side, called Multigaussian LSF hereafter:
\begin{equation}\label{equ:LSFModelButler}
    L_\mathrm{Multi}(x) = a_0 \exp \left(-\frac{x^2}{2 c_0^2}\right) + \sum_{i=1}^{n} a_i \exp \left(-\frac{(x-b_i)^2}{2 c_i^2}\right) \quad \mathrm{.}
\end{equation}
Here, the positions $b_i$ and widths $c_i$ of all Gaussians as well as the central amplitude $a_0$ are usually fixed, leaving $n$ free model parameters, namely the amplitudes of the satellite Gaussians $a_{i\neq0}$. Typically, $n = 10$ offers enough flexibility to account even for small variations and asymmetries. Also note that each LSF model is normalized to unit area, through $L_\mathrm{norm}(x) = L(x) / \sum_{x} L(x)$.

\subsubsection{The continuum \& wavelength models}
\label{subsubsec:ContinuumWavelengthModels}

Both the continuum flux and the wavelength scale of an \'echelle spectrum exhibit complex, large-scale variations along the orders, attributable for example to the blaze-function of the spectrograph, grating anamorphism, optical distortions etc. By modeling the spectrum in individual chunks over rather short pixel ranges, that complexity can be greatly reduced and the continuum flux and wavelengths within a chunk can typically both be described by a simple linear model with a slope $p_\mathrm{slope}$ and an intercept $p_\mathrm{intercept}$.
In the case of the wavelength model $\lambda(x)$, the variable parameters $p_\mathrm{intercept}$ and $p_\mathrm{slope}$ represent the wavelength at the central pixel of the chunk and the dispersion within the chunk, respectively, while for the continuum model $k(x)$ they describe the amplitude (again at the central chunk pixel) and slope of the continuum flux values.

Additionally, we also implemented a quadratic model in \texttt{pyodine}, which can be easily activated and used as an alternative for the wavelength and/or continuum models. In our general adaptation of \texttt{pyodine} to Lick and SONG spectra however, the quadratic model did not show any significant improvements in any of the key result metrics (e.g., chunk velocity scatter, residuals between observed and model spectrum), while increasing the computation time per observation by about 20\% (when using a quadratic instead of linear model both for wavelengths and continuum). In our standard implementation for these two instruments, we therefore keep the linear models as default; for other instruments (or when using larger chunks, compare Section~\ref{subsubsec:SONGSun}), the additional complexity of the quadratic model nevertheless can lead to better results. Also, thanks to the modular design of \texttt{pyodine}, it is straightforward for users to add their own model descriptions if required.

\subsection{The deconvolved stellar template}
\label{subsec:DeconvolvedStellarTemplate}

While the I$_2$ template spectrum can be acquired directly with an FTS, this process is not a viable solution to produce the stellar template, as it would result in too low signal-to-noise ratios $S/N$ for most stars. However, when using the same spectrograph as for the observations, but without the I$_2$ cell in the light path, the stellar spectrum is still affected by the instrumental LSF. \citet{MarcyButler1992} therefore developed a routine where the stellar template spectrum is obtained by ``cleaning'' an I$_2$-free observation spectrum of the star from the LSF in a deconvolution algorithm.

To achieve this, spectra of hot and rapidly rotating stars are obtained with the I$_2$ cell in the light path directly before and after the I$_2$-free stellar observation spectrum. As these stars, typically of spectral class O or B, show essentially no absorption lines in the wavelength range of interest, the recorded spectra only contain the LSF-affected I$_2$ features; that is to say, these stars act as a continuum source with the benefit that the light travels (nearly) the same path through the instrument as for all other observations. When combined and split into chunks, each chunk spectrum $I(x)$ 
can then be modeled by:
\begin{equation}\label{equ:IodineTempModel}
    \hat{I}(x) 
    = k(x) \, T_\mathrm{I2}\big(\lambda(x)\big) * L(x) \quad \mathrm{.}
\end{equation}
Thus we receive a description of the LSF within each chunk, $L(x)$, which we can now use in a deconvolution of the I$_2$-free stellar observation $I_*(x)$: We are interested in the ``true'' stellar object spectrum $T_*(\lambda)$ which satisfies
\begin{equation}\label{equ:Deconvolution1}
    I_*(x) = T_*\big(\lambda(x)\big) * L(x) + N(x) \quad \mathrm{,}
\end{equation}
where $N(x)$ is the noise in the observation. For deconvolving the spectrum, \citet{Butler1996} used a modified version of the Jansson technique which involves an iterative process, following \citet{Gilliland1992}; in \texttt{pyodine}, we incorporate a slightly different recipe developed by \citet{Crilly2002} on the basis of the algorithm by \citet{Agard1989}, which uses the same iterative steps, but pre-filters the input data to accelerate convergence.

The deconvolved stellar template chunks are then saved to file, along with the fitted wavelength parameters of each chunk and the computed barycentric velocity at the time of acquiring the I$_2$-free stellar template spectrum. When modeling the observation spectra as described in Section~\ref{subsec:ModellingObservations}, the chunks in these observations are chosen such that their wavelengths roughly correspond to the template chunk wavelengths after taking the relative barycentric velocities between them into account -- this means that the chunks are co-moving in wavelength space with the barycentric velocity. This ensures that the observation and template chunks will mostly overlap in pixel space, and the only relative shift is due to the true relative RV between both (i.e., corrected for relative barycentric velocity), which is usually small (tens of $\unit[]{m\,s^{-1}}$ for planet-induced RVs, a few $\unit[]{km\,s^{-1}}$ for close stellar binaries).

\begin{figure*}
\centering
\includegraphics[width=\hsize]{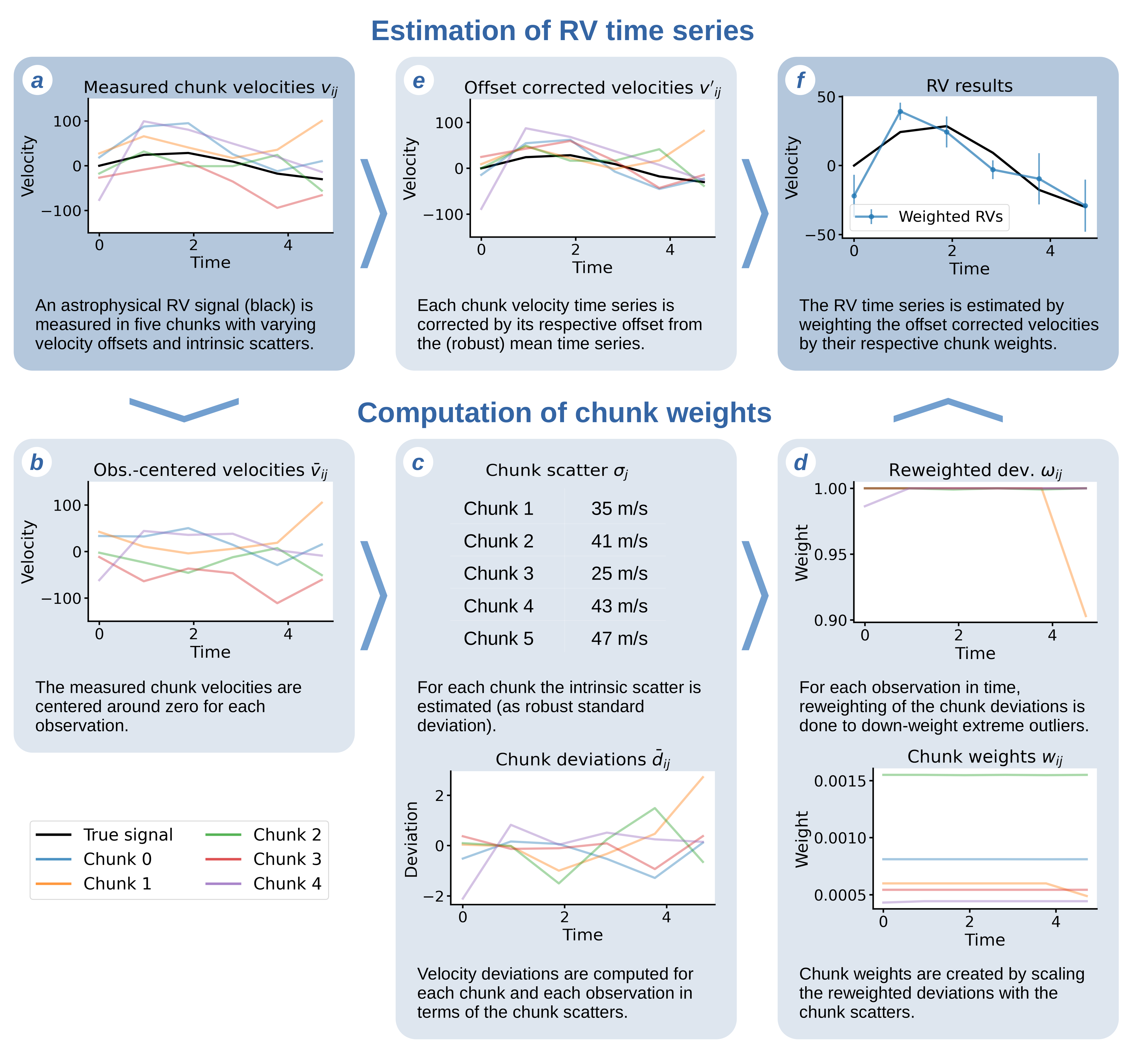}
\caption{Illustration of the combination algorithm of chunk velocities, using a simulated sinusoidal RV signal (black line), ``measured'' at six points in time by five chunks with different noise levels added to the signal (colored lines).}
\label{fig:velocity_comb}
\end{figure*}

\subsection{Combining fitted velocities to an RV time series}
\label{subsec:VelocityCombination}

The modeling process of an observation, as described in Section~\ref{subsec:ModellingObservations}, results in $n_\mathrm{ch}$ individual best-fit velocities $v_{ij}$ for each observation $i$ (where $n_\mathrm{ch}$ is the number of chunks within one spectrum). Since the chunks do not overlap, all these velocities can be treated as independent estimates of the true RV of the star at the time of the observation, and a simple mean or median over all $v_j$ would be the easiest way to compute an overall RV estimate. However, as \citet{Butler1996} discuss in detail, different chunks contain vastly different Doppler information because the number and depth of stellar absorption lines within each chunk varies greatly. This leads to some chunks contributing a better estimate of the stellar RV, and the overall RV determination will thus greatly improve if we assign a larger weight to these chunks. Furthermore, by weighting chunks differently we can weaken the contribution of chunks that are corrupted (e.g., through instrumental defects).

The RV time series of a star is therefore computed taking all chunk velocities from all $n_\mathrm{obs}$ observations $v_{ij}$ into account, which is a 2D-matrix of shape $(n_\mathrm{obs},n_\mathrm{ch})$. Fig.~\ref{fig:velocity_comb} illustrates the velocity combination algorithm for a simulated ``true'' signal and ``measured'' chunk velocities (six observations, five chunks with different noise properties, see box~a in Fig.~\ref{fig:velocity_comb}). Within each observation, the velocities are first centered around $0$ by subtracting an outlier-resistant mean\footnote{The outlier-resistant (robust) mean and standard deviation are computed using a Python implementation of the \texttt{IDL ROBUST} Package: \url{https://idlastro.gsfc.nasa.gov/contents.html#C17}} of $v_{ij}$ over the chunks $j$, $\mathrm{mean}(v_{ij}, j)$ (Fig.~\ref{fig:velocity_comb}b):
\begin{equation}\label{equ:VelocityComb1}
   \bar{v}_{ij} = v_{ij} - \mathrm{mean}(v_{ij}, j) \quad \mathrm{.}
\end{equation}
From the centered velocities $\bar{v}_{ij}$, we can now compute the scatter within each chunk time series $\sigma_j$ by taking the robust standard deviation $\mathrm{std}()$ over observations $i$ (Fig.~\ref{fig:velocity_comb}c, top):
\begin{equation}\label{equ:VelocityComb2}
   \sigma_j = \mathrm{std}(\bar{v}_{ij}, i) \quad \mathrm{.}
\end{equation}
Now, for each chunk we can compute the deviation $\delta_{ij}$ from the mean of its chunk time series (i.e., over all observations) in terms of the overall scatter within that chunk time series $\sigma_j$. These deviations are then again centered around $0$ by subtracting the mean of all deviations within an observation (Fig.~\ref{fig:velocity_comb}c, bottom):
\begin{equation}\label{equ:VelocityComb3}
   d_{ij} = \frac{\bar{v}_{ij} - \mathrm{mean}(\bar{v}_{ij}, i)}{\sigma_j} \quad \mathrm{,} \quad \bar{d}_{ij} = d_{ij} - \mathrm{mean}(d_{ij}, j) \quad \mathrm{.}
\end{equation}
The chunk deviations $\bar{d}_{ij}$ can be used to compute weights for the chunks. To increase the contrast between good chunks (small absolute $\bar{d}_{ij}$) and bad chunks (large absolute $\bar{d}_{ij}$), we use a re-weight function, inspired by \citet{stetson1989techniques}, of the form:
\begin{equation}\label{equ:VelocityComb4}
    \omega_{ij} = \frac{1}{1 + \left( |\bar{d}_{ij}| / (a \cdot s) \right)^\beta} \quad \mathrm{,}
\end{equation}
where the constants $a$, $s$ and $\beta$ can be set by the user in a configuration file \citep[compare Chapter 3.2 in][for a detailed explanation of these constants]{stetson1989techniques}. We found that $a=1.8$, $s=2$ and $\beta=8$ deliver good results. The resulting weights $\omega_{ij}$ are distributed in the interval $[0,1]$, where chunks with small absolute $\bar{d}_{ij}$ are close to $1$ (see Fig.~\ref{fig:velocity_comb}d, top). To arrive at the final weights for the velocity combination $w_{ij}$, we scale the $\omega_{ij}$ by the chunk time series scatter $\sigma_j$ (Fig.~\ref{fig:velocity_comb}d, bottom):
\begin{equation}\label{equ:VelocityComb5}
   w_{ij} = \left( \frac{\omega_{ij}}{\sigma_j} \right)^2 \quad \mathrm{.}
\end{equation}
Before we weight the chunk velocities, we aim to center each chunk time series around $0$; therefore we compute chunk time series offsets from the observation means $c_j$, and again center them around $0$ to arrive at $\bar{c}_j$:
\begin{equation}\label{equ:VelocityComb6}
   c_j = \mathrm{mean}(\bar{v}_{ij}, i) \quad \mathrm{,} \quad \bar{c}_j = c_j - \mathrm{mean}(c_j, j) \quad \mathrm{.}
\end{equation}
Now we can correct the original chunk velocities $v_{ij}$ by the chunk time series offsets $\bar{c}_j$ (Fig.~\ref{fig:velocity_comb}e):
\begin{equation}\label{equ:VelocityComb7}
   v_{ij}' = v_{ij} - \bar{c}_j \quad \mathrm{.}
\end{equation}
Finally, we use these corrected chunk velocities to compute the weighted mean for each observation, which results in the RV time series of the star:
\begin{equation}\label{equ:VelocityComb8}
    RV_i = \frac{\sum_j v_{ij}' w_{ij}}{\sum_j w_{ij}} \quad \mathrm{.}
\end{equation}
For each observation, we determine an uncertainty estimate according to:
\begin{equation}\label{equ:VelocityComb9}
    \sigma_i^\mathrm{RV} = \sqrt{\frac{\sum_j w_{ij} \cdot (RV_i - v_{ij}')^2}{(N_i - 1) \cdot \sum_j w_{ij}}} \quad \mathrm{,}
\end{equation}
where $N_i$ is the number of chunks with good velocities (non-NaN, i.e., the fit converged successfully) in observation $i$. Fig.~\ref{fig:velocity_comb}f shows the RV time series for the simulated velocities.

The RV estimate $RV_i$ in Equation~\ref{equ:VelocityComb8} contains contributions both from the observed star as well as the barycentric motion of the observatory along the line-of-sight. To arrive at the final RV time series with only stellar velocity contributions $RV_i^*$, we use the open-source Python package \texttt{barycorrpy} \citep{Kanodia2018}. The package follows the routines outlined in \citet{Wright2014} to correct measured redshifts $z_i^\mathrm{meas}$ for barycentric velocities at the $\unit[1]{cm\,s^{-1}}$-level:
\begin{equation}\label{equ:VelocityComb10}
    RV_i^* = z_i^* \cdot c = \mathrm{bary}(z_i^\mathrm{meas}) \cdot c \quad \mathrm{.}
\end{equation}
We note that \textbf{absolute} measured redshifts are required for the barycentric correction to work at highest precision \citep[compare Equation~10 in][]{Wright2014}. In \texttt{pyodine}, the RV estimates $RV_i$ are relative to the template observation, which is usually affected by some non-zero Doppler shift with respect to laboratory wavelengths itself. Therefore, we allow to incorporate a constant RV offset $RV_\mathrm{off}$ (usually the template velocity offset, estimated through cross-correlation of the stellar template with a reference spectrum, see Section~\ref{subsec:workflow}), to arrive at the absolute measured redshifts:
\begin{equation}\label{equ:VelocityComb11}
    z_i^\mathrm{meas} = \frac{RV_i + RV_\mathrm{off}}{c} \quad \mathrm{.}
\end{equation}
Be aware that the template velocity offset is only accurate on the order of a few $\unit[100]{m\,s^{-1}}$, as it is determined through a simple cross-correlation. However, the method outlined above suffices to achieve a barycentric correction better than $\unit[1]{m\,s^{-1}}$ in all cases, and much better in most.

\section{Code implementation}
\label{sec:CodeImplementation}

\subsection{Core modules}

The mathematical concepts described above are implemented within the \texttt{pyodine} software package, which is completely written in Python~3 and available for download and use under MIT license\footnote{\url{https://mit-license.org/}}. 
All components of the algorithm (including models, sub-models, observations, template spectra, and routines for chunk-definition, deconvolution, and fitting) are represented as Python objects, encapsulating for instance instrument-specific code within each instrument class. The philosophy of \texttt{pyodine}, as a framework for I$_2$ reduction, is that all these components should be easily swappable.
Some open-source Python packages are used for specific tasks, for example \texttt{lmfit} for fitting and \texttt{barycorrpy} \citep[][]{Kanodia2018} for the computation of barycentric velocities. Additionally, \texttt{pyodine} comes with a number of convenience functions for the creation of analysis plots or loading and saving of results.

Instrument- and user-specific code is largely bundled in separate directories, called \texttt{utilities} modules hereafter. These contain functions which define how to load input spectra and required meta-data, for instance times of observations or name of the observed object, which differs from instrument to instrument. Similarly, some fixed parameter values for the I$_2$ cell code may differ for different instruments, such as the pixel width of chunks or the positions of satellite Gaussians in the Multigaussian LSF (Equation~\ref{equ:LSFModelButler}); for a given instrument, all these parameters are therefore defined in parameter input files which reside inside the instrument's \texttt{utilities} module.

\begin{figure*}
\centering
\includegraphics[width=\hsize]{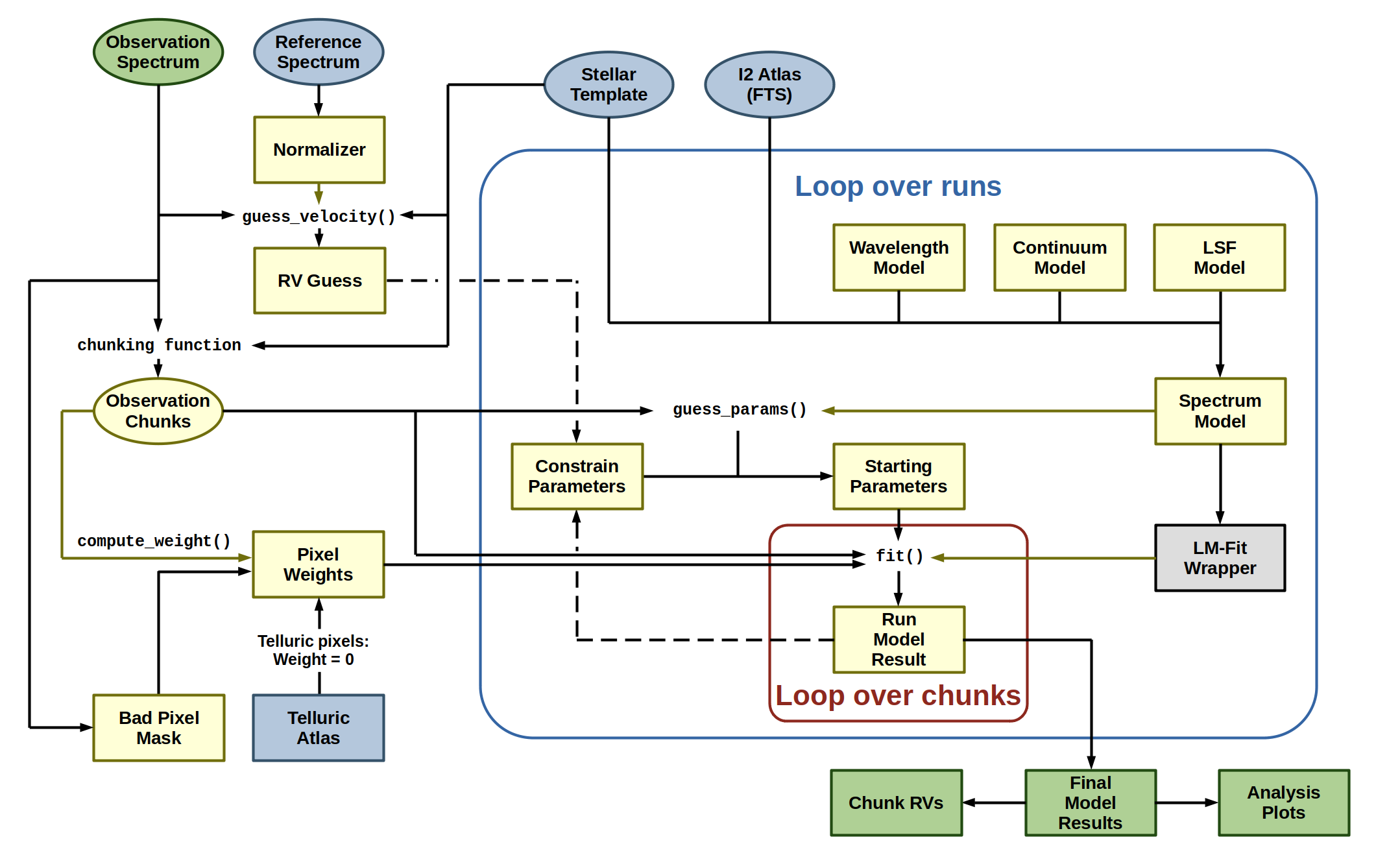}
\caption{Flowchart of the I$_2$ analysis of an observation.}
\label{fig:pyodineflowchart}
\end{figure*}

So far, we have adapted \texttt{pyodine} to work with spectra from two instruments, the SONG Hertzsprung spectrograph in Tenerife/Spain and the Hamilton spectrograph at Lick observatory/USA. For each of these instruments a \texttt{utilities} module exists with all required parameters and functions (called \texttt{utilities\_song} and \texttt{utilities\_lick}, respectively), and a third module \texttt{utilities\_waltz} has been set up for the so-far untested new spectrograph at the Waltz telescope. In Section~\ref{sec:Results}, we briefly describe the most important parameters employed and present examples of test results on data from these two instruments.

\subsection{Typical workflow}
\label{subsec:workflow}

Following the philosophy of \texttt{pyodine}, the main routines containing the workflows for template creation, observation modeling, and the combination and weighting of chunk velocities are generalized for all instruments; all differences between instruments that affect the major workflow can be controlled through the parameters in the \texttt{utilities} modules.

Fig.~\ref{fig:pyodineflowchart} schematically presents the main steps of the workflow for modeling an observation, and we briefly explain it here:

\begin{enumerate}
    \item The observed spectrum and corresponding stellar template ($I_\mathrm{obs}$ and $T_*$) are loaded from the disk.
    
    \item A \texttt{Normalizer} object is initialized with a high-$S/N$ reference spectrum, and a first RV guess of the observation relative to the template is computed by cross-correlating each of the two with the reference spectrum (using 
    orders that lie outside the I$_2$-affected wavelength region).
    At the moment \texttt{pyodine} comes with atlases of Arcturus and the Sun \citep{Hinkle2000} as reference spectra, but users can also add their own choices (through minimum changes to the code and the respective \texttt{utilities} modules).
    
    \item The observation is split into chunks that roughly correspond to the template chunks in wavelength space (after correcting for the relative barycentric velocity shift between both, see Section~\ref{subsec:DeconvolvedStellarTemplate}), and pixel weights are computed -- in this step information from a bad pixel mask or a telluric atlas can optionally be used.
    
    \item The modeling is performed, optionally in multiple runs, where the exact choice of (sub-)models can be varied between runs, and parameter results from one modeling run can be used to change starting values or constraints of parameters in a later one. Each run works as follows:
    
    \begin{enumerate}
        \item Set up the spectrum model (with the sub-models desired for this run), and use it to initialize the fitter object.
        
        \item Choose appropriate starting values and possibly constraints for the model parameters. If desired, fit results from an earlier run can be used here.
        
        \item Loop over the spectrum chunks and model each one. Store the fit results in an internal data object.
        
        \item Possibly save this run's relevant model information and fit results for all chunks to file, as well as automatically created analysis plots.
    \end{enumerate}
    
    For example, in our tests on SONG and Lick spectra, a modeling procedure in two runs proved to deliver good results, where a Singlegaussian LSF model is used in the first run and the results serve as starting guesses for a second run with a Multigaussian LSF. The number of runs and their model and parameter setup are controlled through the parameter files in the \texttt{utilities} modules.
\end{enumerate}

From the saved fit results of a number of observations, the velocities can later be loaded and combined to a RV time series using a routine that follows the steps described in Section~\ref{subsec:VelocityCombination}.

\begin{figure*}
\centering
\includegraphics[width=\hsize]{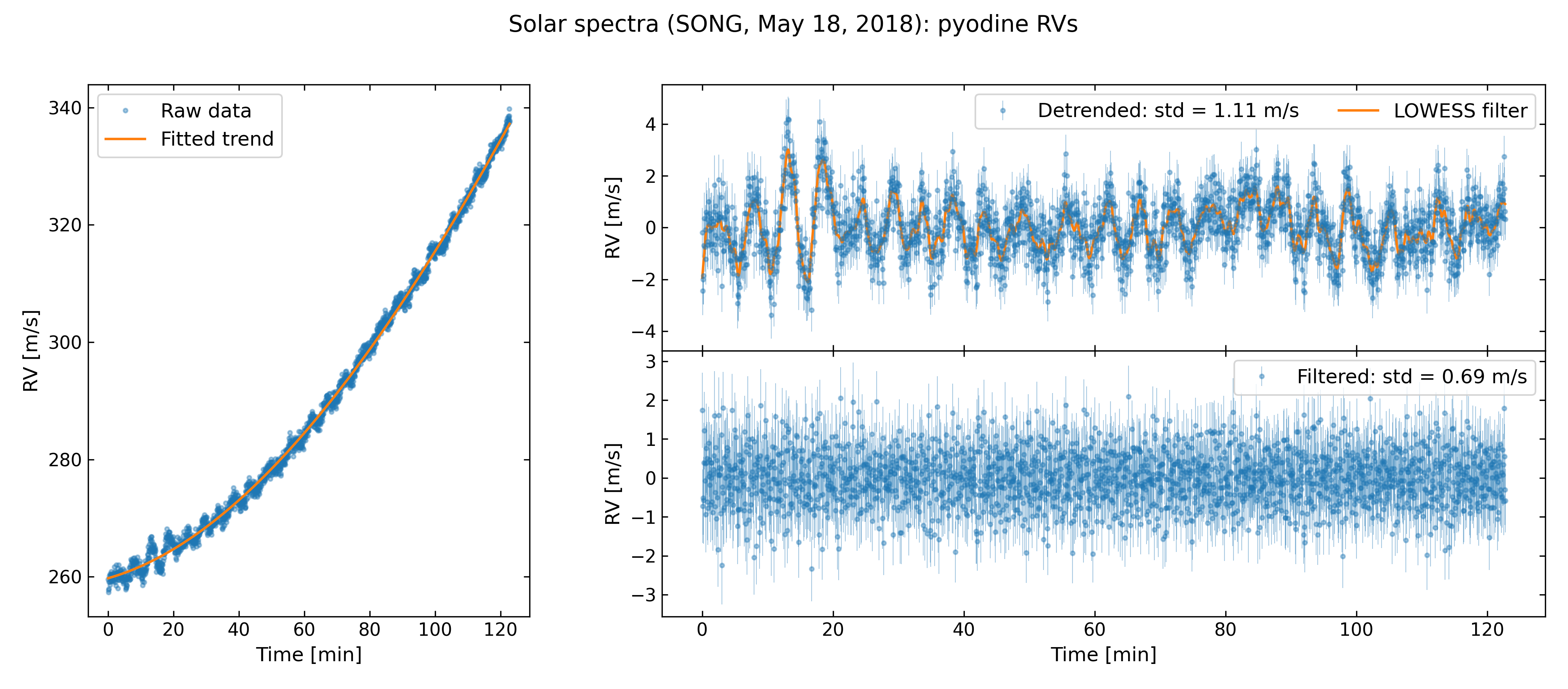}
\caption{RV time series of 2000 solar spectra obtained with Solar-SONG, and modeled and computed by \texttt{pyodine}. \textit{Left:} The measured RVs along with a fitted trend to correct for relative movement of the observatory with respect to the sun. \textit{Right:} Detrended RVs along with a fitted LOWESS filter (window size $0.8\%$ of the whole time series) to correct for the solar p-mode oscillations (top), and smoothed RVs (bottom).}
\label{fig:SONGsunRVs}
\end{figure*}

In the process of creating a deconvolved stellar template, the modeling of the hot-star observation works largely as outlined above, with a few differences:

\begin{itemize}
    \item No stellar template is required, and no prior RV guess is computed.
    
    \item The chunks can be freely defined, and the definition chosen here reflects on the chunk properties of the deconvolved stellar template and the later observations, as they all correspond in (barycentric-corrected) wavelength space. \texttt{pyodine} offers different chunking algorithms for the user: One can either pack as many chunks of a given size as possible into the orders, thus making full use of the accessible spectral range -- this should normally be used for precise RVs. However, we also offer the possibility of only chunking specific wavelength regions -- the user can thus track velocity variations in specific absorption lines for example, which might prove valuable in asteroseismic measurements (compare Section~\ref{subsubsec:SONGSun}).
    
    \item After the last run (item~4. in above flow outline), the best-fit results of the hot-star model are used to divide into chunks and deconvolve the input stellar template observation ($I_*$), and the resulting deconvolved template chunks are then saved to file.
\end{itemize}

The code was tested extensively on a desktop PC with a $\unit[3.1]{GHz}$ processor, and the workflow described above required typical computation times of $2.5$ to $\unit[3]{min}$ for the modeling of a single SONG observation, and $4$ to $\unit[5]{min}$ for a Lick spectrum. To accelerate the analysis of large time series of observations, \texttt{pyodine} offers the possibility to parallelize the modeling of spectra; this is currently done using the Python package \texttt{pathos} \citep{McKerns2011}. In our tests we ran the software on 12 cores of a desktop computer, and the reduction of the overall 2061 SONG observations of $\sigma$~Draconis presented in Section~\ref{subsec:SONGSpectra} thus took $\unit[11.1]{hr}$, while the Lick time series of 91 spectra of HIP~36616 (Section~\ref{subsec:LickSpectra}) was modeled in $\unit[33]{min}$.
Note that the subsequent combination and weighting of the chunk velocities of all modeled observations, as outlined in Section~\ref{subsec:VelocityCombination}, computes within a few seconds (for short time series) to some minutes (for very long time series of $\mathcal{O}(10^4)$ observations).\footnote{The bottleneck here is not the actual computation itself, but the reading of the model results from the disk.}

\section{First results}
\label{sec:Results}

\subsection{SONG spectra}
\label{subsec:SONGSpectra}

The SONG instrument is a high-resolution, cross-dispersed \'echelle spectrograph, fed through the coud\'e path by the 1\,m robotic Hertzsprung telescope at Teide Observatory on the island of Tenerife, Spain \citep{Andersen2014,Andersen2019}. It was built as part of an initiative to set up a world-wide network of spectrographs dedicated to asteroseismic measurements through RVs \citep{Grundahl2007}, and has been in operation since 2014. With the typically used slit the spectrograph reaches a resolving power of $\sim 90,000$, with a maximum of $\sim 115,000$ for the narrowest slit.

The spectral orders are extracted and wavelength-calibrated using a code based on the IDL package \texttt{REDUCE} \citep{Piskunov2002}.
RVs are computed through the I$_2$ cell method with the IDL code \texttt{iSONG}, and for that 24 \'echelle orders of 2048 pixels each are used, covering the wavelength range from roughly $4994$ to $\unit[6208]{\AA}$. They are split into 22 chunks of 91 pixels in width per order, resulting in a total of 528 chunks for each spectrum. Typically the Multigaussian model is used for fitting the LSF (Equation~\ref{equ:LSFModelButler}), and in the construction of the model an oversampling factor of 6 is employed. 

To evaluate the performance of \texttt{pyodine} and compare it to the \texttt{iSONG} results, we tested our code on SONG observations of several different targets, while using exactly the same parameter setup as \texttt{iSONG}. The following two sections show results from a series of solar observations, which test the internal precision of the software, and from observations of the star $\sigma$~Draconis, which illustrate an instrumental artifact of SONG.

\begin{figure*}
\centering
\includegraphics[width=\hsize]{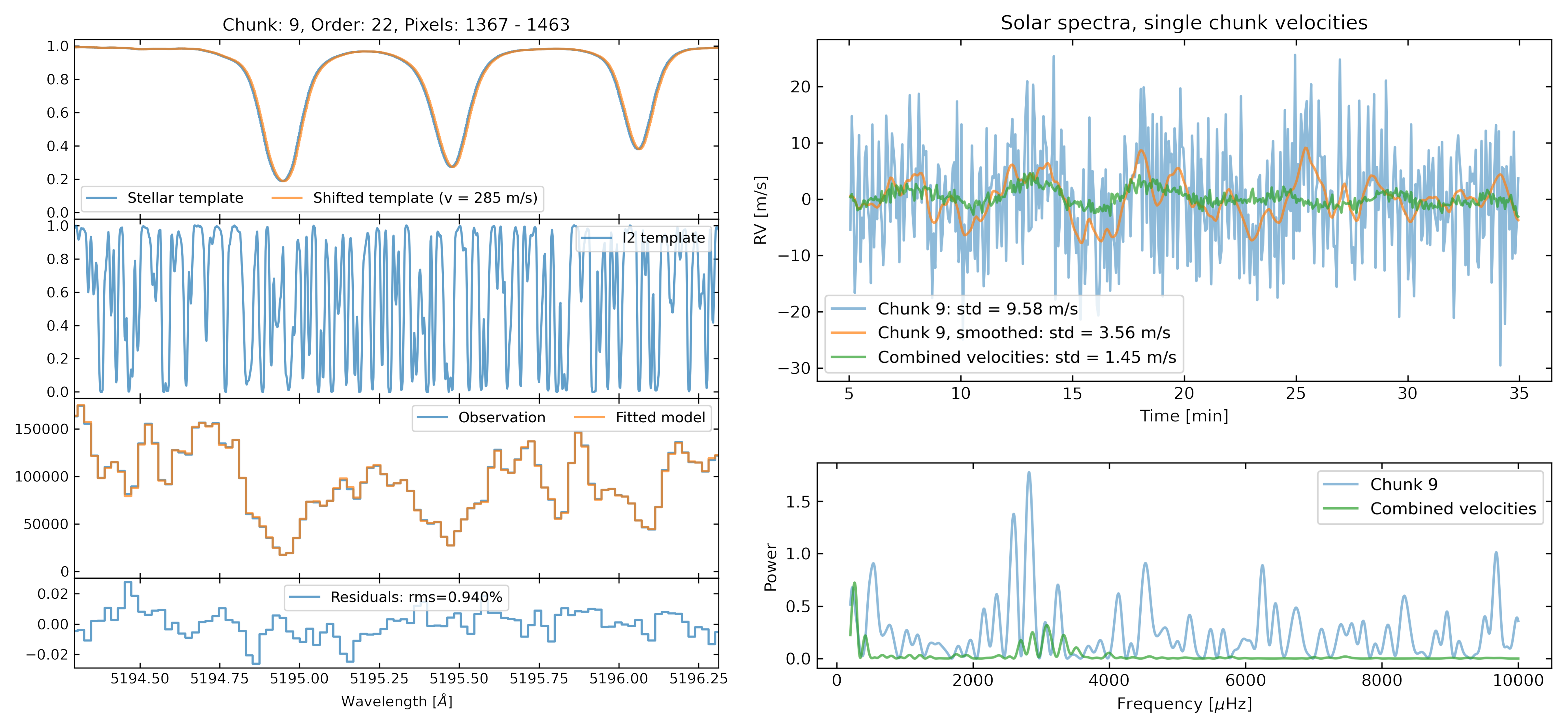}
\caption{Results for a single chunk (chunk index 9) of a Solar-SONG observation, centered around three Fe~I-lines around $\unit[5195.5]{\AA}$. \textit{Left:} Spectrum and best-fit model components. \textit{Right top:} Zoom-in on a $\unit[30]{min}$ window of the chunk's best-fit velocity time series (blue), along with a smoothed time series (orange), and the detrended combined RVs as in Fig.~\ref{fig:SONGsunRVs} (green). \textit{Right bottom:} Power spectra of the full $\unit[2]{hr}$ time series of chunk-9 velocities (blue) and the combined detrended RVs (green).}
\label{fig:SONGsunsinglechunk}
\end{figure*}

\subsubsection{Solar data}
\label{subsubsec:SONGSun}

For the Solar-SONG initiative, the SONG spectrograph is equipped with an optical multi-mode fiber whose one end is mounted on a solar tracker outside the telescope dome, while the other end is focused on the entrance slit of the spectrograph \citep{Palle2013,Andersen2019sun,Breton2022}. SONG thus allows to take precise RV data of the Sun during daytime, which contributes to a deeper understanding of the solar interior.

Additionally, due to the extraordinary high $S/N$ that is achieved in observations of the Sun even with short exposure times of $\unit[1]{s}$ and less, the short-term RV precision of the instrument and reduction code can be tested on time series of very high-cadence spectra. To probe the precision of \texttt{pyodine} and compare it to the results produced by the dedicated reduction pipeline \texttt{iSONG}, we modeled a series of $2000$ spectra of the Sun obtained over the course of $\unit[2]{hr}$ on May 18, 2018 (on average more than 16 spectra per minute). As stellar template, we used a spectral atlas of the Sun \citep{Hinkle2000}, which was transformed into the required format.

The fitted chunk velocities were then combined as described in Section~\ref{subsec:VelocityCombination}, but without correcting for barycentric movement, to create the RV time series shown in the left plot of Fig.~\ref{fig:SONGsunRVs}. A long trend is seen, caused by the Earth's rotation, which we fit with a \nth{3}-degree polynomial to remove it from the data. In the detrended time series (top right graph of Fig.~\ref{fig:SONGsunRVs}) oscillations induced by solar p-mode pulsations are clearly visible with a peak-to-peak amplitude around $\unit[4]{m \,s^{-1}}$, leading to a RV scatter (standard deviation) of $\unit[1.11]{m \,s^{-1}}$. Qualitatively these results are well in line with an analysis performed by \citet{Andersen2019sun} on a much larger data set of Solar-SONG observations \citep[compare Fig.~2 in][]{Andersen2019sun}.

To exclude the systematics of the p-mode pulsations and get to the bottom of the intrinsic RV precision, we used a LOWESS filter \citep[locally weighted regression and smoothing;][]{Cleveland1979} to smooth the data. We tested different filter window sizes between $0.4$ and $2.4\%$ of the whole data set, corresponding to time windows between $0.5$ and $\unit[3]{min}$, which reduced the standard deviation of the filtered RV time series to values between $0.64$ and $\unit[0.77]{m\,s^{-1}}$. For the bottom right graph of Fig.~\ref{fig:SONGsunRVs}, we chose to adopt a window size of $0.8\%$ (roughly $\unit[1]{min}$), which is short enough to completely filter out the dominant solar p-mode pulsations with periods between $5$ and $\unit[6]{min}$, but still averages over sufficiently many observations to probe the point-to-point scatter. The filter reduces the standard deviation of the RVs to $\unit[0.69]{m \,s^{-1}}$, which is a little smaller than the mean of the individual RV uncertainties as computed by \texttt{pyodine}, $\bar{\sigma}_\mathrm{RV} = \unit[0.81 \pm 0.07]{m \,s^{-1}}$.

The results from \texttt{pyodine} for this $\unit[2]{hr}$-time series of solar spectra match the original RVs produced by the dedicated SONG pipeline nearly perfectly: When performing the same analysis steps as presented above on the \texttt{iSONG} RVs, the detrended time series scatters with $\unit[1.12]{m \,s^{-1}}$, and the LOWESS-filtered time series (window size of $0.8\%$) with $\unit[0.69]{m \,s^{-1}}$ (see Fig.~\ref{fig:SONGsunRVs_iSONG} in Appendix~\ref{app:1} for the respective results plots).


We furthermore used the same Solar-SONG observations to test the flexibility of \texttt{pyodine} with respect to chunking and more complex modeling: By providing appropriate start and stop wavelengths, we created 28 chunks centered around prominent absorption features, such as the Na~$\mathrm{D}_1+\mathrm{D}_2$ and several strong Fe, Mg, Mn and Ti lines. These chunks have varying widths, with those around broader lines being much wider than the usual 91 pixels / $\unit[2]{\AA}$ in order to still include sufficient parts of the continuum; therefore, when using these chunks to model the same 2000 solar spectra as described above, we adopted quadratic wavelength and continuum models to account for non-linearities in the wavelength scale and blaze function.

Fig.~\ref{fig:SONGsunsinglechunk}, left, shows the data and model results of chunk~9, centered around three closely spaced Fe~I-lines around $\unit[5195.5]{\AA}$, for one solar observation. The relative (outlier-resistant) residuals between data and model have an rms of $0.94\%$, which corresponds to the median of this chunk's residuals over all observations. While the width of this chunk has been chosen quite conservatively to be nearly exactly $\unit[2]{\AA}$, some wider chunks (up to $\unit[5]{\AA}$ in width) centered around other absorption lines show similar residuals, indicating the good performance of the non-linear wavelength and continuum models.

The top right-hand graph in Fig.~\ref{fig:SONGsunsinglechunk} shows a $\unit[30]{min}$ window of the best-fit velocity time series of chunk~9, along with a smoothed version of the data (using the same LOWESS filter as described above), and the detrended combined RVs from above. The single-chunk velocities vary much more than the combined RVs and their point-to-point scatter is much larger
. Accordingly, the power spectrum of the chunk velocities of all 2000 observations (Fig.~\ref{fig:SONGsunsinglechunk}, right bottom, in blue) shows a higher noise floor compared to the one 
of the combined RVs (same graph, in green), 
with a visible power excess around the p-mode induced signal frequencies between roughly $2500$ and $\unit[3700]{\si{\micro}\si{Hz}}$ visible in the combined RVs \citep[compare also Fig.~3 in][]{Andersen2019}. This is the case for most of our tested chunks centered around single absorption lines. 
We expect that the noise can be reduced significantly by including more observations; a similar analysis like this performed on a larger data set can then yield additional scientific value to a simple RV time series of combined chunk velocities (see Section~\ref{sec:Conclusions} for further discussion).

\begin{figure}
\centering
\includegraphics[width=\hsize]{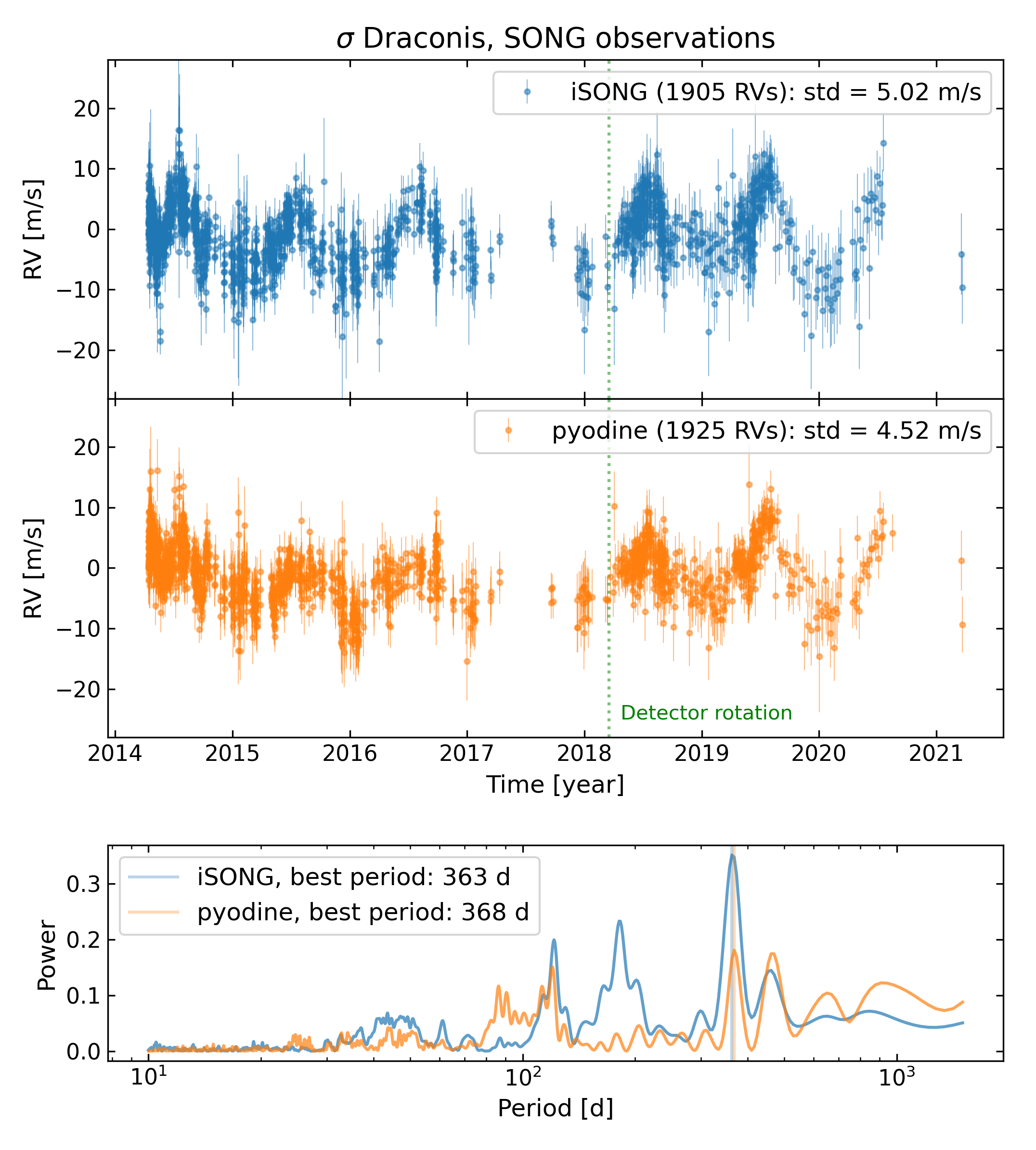}
\caption{RV results for $\sigma$~Dra, from SONG spectra. \textit{Top plots:} RV time series computed with \texttt{iSONG} (top) and \texttt{pyodine} (bottom); the green-dotted line marks the date of the detector rotation. \textit{Bottom plot:} GLS periodograms of the \texttt{iSONG} and \texttt{pyodine} time series, respectively.}
\label{fig:SONGsigdraRVs}
\end{figure}

\subsubsection{Sigma Draconis}
\label{subsubsec:SONGSigDra}

The main-sequence star $\sigma$~Draconis ($\sigma$~Dra, HD~185144) has long served as RV standard star \citep[spectral type K0V,][]{Keenan1989}, and has been observed extensively by the SONG telescope since the beginning of operations in April 2014: Until early October 2021, a total of 2061 spectra were obtained with the standard I$_2$ cell to monitor the long-term precision of the instrument. The star has a V magnitude of $4.68$ \citep{Oja1984}, and typical exposure times with SONG were $5$ -- $\unit[10]{min}$.

The RV time series of these observations, as computed by the \texttt{iSONG} pipeline and shown in the top-most plot of Fig.~\ref{fig:SONGsigdraRVs}, is modulated by a signal of roughly yearly period and amplitude of $\sim\unit[20]{m \,s^{-1}}$, leading to a scatter of roughly $\unit[5]{m\,s^{-1}}$. A generalized Lomb-Scargle periodogram \citep[GLS,][]{Zechmeister2009} of the data shows a strong peak around a period of \unit[1]{yr}, along with additional smaller peaks at roughly half and one-third of that period probably caused by harmonics of the $\unit[1]{yr}$-signal (blue line in Fig.~\ref{fig:SONGsigdraRVs}, bottom). Similarly, the RVs computed with \texttt{pyodine} show clear variations, albeit smaller in amplitude (overall scatter $\unit[4.5]{m\,s^{-1}}$), especially in some parts of the time series (e.g., between  
2015 and mid-2017, see Fig.~\ref{fig:SONGsigdraRVs} second plot from top). Consequently, the GLS periodogram of the \texttt{pyodine} RVs displays less prominent peaks, particularly around the yearly period, and the half-year period even completely vanished (orange line in Fig.~\ref{fig:SONGsigdraRVs}, bottom).

The cause of these RV variations are not completely clear at the moment. While \citet{Vogt2014} detect a potential planet around $\sigma$~Dra with a $\unit[308]{d}$-period, they report a small RV amplitude $< \unit[2]{m \,s^{-1}}$; it is thus implausible that a planetary modulation is the cause for the observed SONG RV variations. Furthermore, we have observed similar modulations with $\unit[1]{yr}$-periods (and higher frequencies as well) in other targets observed over long time-scales with SONG, and thus we ascribe these variations to an instrumental effect. However, at present we do not fully understand the origin of this as our efforts so far have not yielded a conclusive cause. We have confirmed that our barycentric correction works correctly, and for the star $\sigma$~Dra we obtain a similar result when using two different I$_2$ cells. Paul Butler has kindly performed a completely independent reduction (including spectral extraction) of the data and also finds a similar modulation. Furthermore, on March 18, 2018, we have attempted a $90^\circ$ rotation of the detector (marked by the green-dotted line in Fig.~\ref{fig:SONGsigdraRVs}), with no effect.

The results presented in Fig.~\ref{fig:SONGsigdraRVs} suggest that there is some dependence on the software used as the modulation seen in the \texttt{pyodine} time series is less pronounced -- although they are still clearly visible. Our present working hypothesis is that the root cause comes from a low-level problem in the detector readout, and the $\unit[1]{yr}$ modulation as well as shorter-period signals originate from the time-varying blending of the stellar and I$_2$ absorption lines due to the orbital motion of the Earth.

\subsection{Lick spectra: HIP~36616}
\label{subsec:LickSpectra}

The Hamilton spectrograph at Lick observatory was one of the first instruments used in the search of extrasolar planets, and it served as the testbed in the development of the I$_2$ cell method \citep{Vogt1987,Butler1996}. It is a cross-dispersed \'echelle spectrograph with a resolving power of $62,000$, fed by both the 3m Shane telescope and the 0.6m Coud\'e Auxiliary Telescope (CAT). When used for RV measurements, the \texttt{Butler} code served as the standard reduction pipeline.

From an RV survey of G- and K-giant stars conducted at Lick observatory between 1999 and 2012, we possess a large database of reduced spectra obtained with the combination of the Hamilton spectrograph and CAT, and their RVs were determined through the \texttt{Butler} code \citep[see e.g.,][]{Frink2001,Reffert2015}. This data allowed the detection of several (multi-)planetary and stellar binary systems, amongst them the first discovery of an exoplanet orbiting an evolved star \citep[$\iota$~Dra,][]{Frink2002}.

In the development of \texttt{pyodine}, we made use of these data to test the performance of the software and compare it to the results from the \texttt{Butler} code. Using information from \cite{Butler1996} and result products in our data base, we aimed to reconstruct the parameters employed in the \texttt{Butler} code and setup our software accordingly: We use 16 orders of 1851 pixels each, covering a total wavelength range between roughly $5016$ and $\unit[5872]{\AA}$. Each order is split into 44 chunks of width 40 pixels, resulting in a total of 704 chunks over all orders. The model is constructed with an oversampling factor of 4, and using the Multigaussian LSF model in the final run.

Overall, \texttt{pyodine} seems to perform similarly as the \texttt{Butler} code, as for a number of targets no significant differences could be registered between the RV time series from the two reductions. In the following, we present some examples of results for observations of one of our targets, HIP~36616, where some small differences are visible.

\begin{figure}
\centering
\includegraphics[width=\hsize]{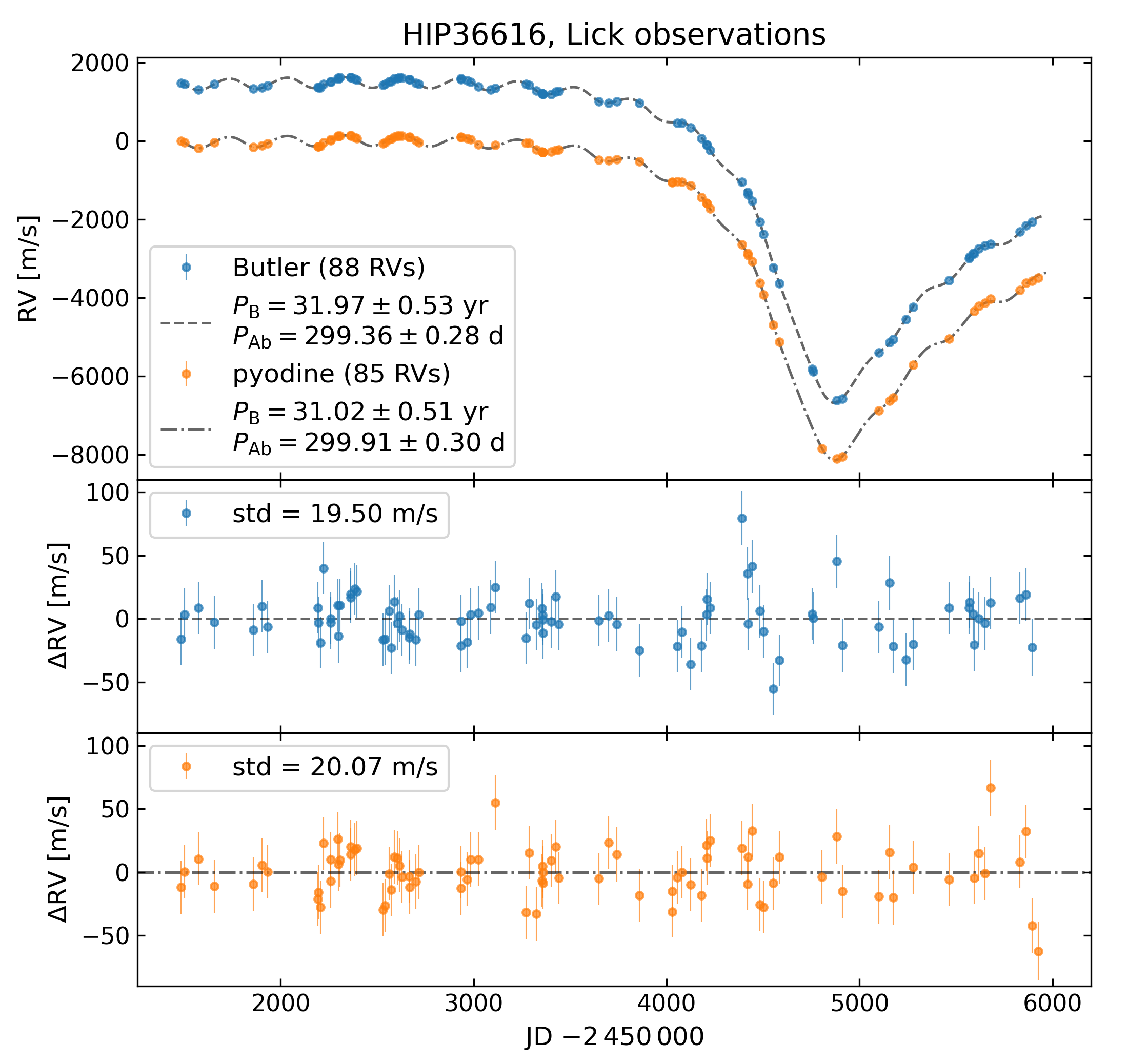}
\caption{RV results of the star HIP~36616, extracted from spectra recorded at Lick Observatory. \textit{Top:} RV time series as computed with the \texttt{Butler} code (blue data points) and \texttt{pyodine} (orange). The dashed and dashed-dotted lines are the best double-Keplerian fits to the \texttt{Butler} and \texttt{pyodine} RVs, respectively. \textit{Middle and bottom:} Residuals between the RVs and best fits for the \texttt{Butler} and \texttt{pyodine} time series, respectively.}
\label{fig:LickHIP36616RVs}
\end{figure}

HIP~36616 (HD~59686) is a single-lined stellar binary system, with the main component HIP~36616~A being a horizontal-branch (HB) star of spectral type K2~III \citep{Reffert2015} and $V = \unit[5.45]{mag}$ \citep{vanLeeuwen2007}. It is part of our Lick RV survey of evolved stars, and based on 88 RVs \citet{Ortiz2016} constrained the orbit of the invisible stellar companion HIP~36616~B and discovered RV variations indicative of a planetary companion orbiting the main component. In a dynamical analysis \citet{Trifonov2018} later showed that despite the high eccentricity of the stellar binary ($e_\mathrm{B} \sim 0.73$) long-term stable orbits of the proposed planet exist.

We used the overall 91 I$_2$ observations of HIP~36616 in our data base, covering the years 1999 until 2011, to test \texttt{pyodine} on Lick spectra and compare the resulting RVs to the original ones determined through the \texttt{Butler} code. Fig.~\ref{fig:LickHIP36616RVs}, top-most graph, shows the RV time series as determined by \texttt{Butler} from the Lick observations (blue data points), and the \texttt{pyodine} RVs computed from the same spectra (orange, offset by $\unit[-1500]{m \,s^{-1}}$). The \texttt{pyodine} data set consists of 85 RVs, three less than the \texttt{Butler} reduction, as the modeling failed for some observations; this is at least partly caused by the fact that the extracted Lick spectra in our data base are not wavelength calibrated yet, and our construction of wavelength solutions \citep[using a self-built pipeline based on CERES routines, see][]{Brahm2017} returned bad results in some cases due to low quality of the respective ThAr calibration spectra. However, \texttt{pyodine} also successfully reduced three observations for which we do not possess any \texttt{Butler} RVs.

\begin{table}

    \centering
    \caption{Orbital parameters of the best-fit Keplerian model to the combined \texttt{pyodine} Lick and SONG RVs of the HIP~36616 system.}
    \label{table:HIP36616}

    \begin{tabular}{lrrrrrrrr}     

    \hline\hline  \noalign{\vskip 0.7mm}
    Parameter \hspace{0.0 mm}& HIP~36616~Ab & HIP~36616~B \\
    \hline \noalign{\vskip 0.7mm}

        $K$ [m\,s$^{-1}$]             &    131.20 $\pm$      2.03 &   4009.29 $\pm$      5.54 \\
        $P$ [day]                     &    300.01 $\pm$      0.08 &  11256.53 $\pm$    111.40 \\
        $e$                           &      0.05 $\pm$      0.01 &      0.72 $\pm$      $2 \cdot 10^{-3}$ \\
        $\omega$ [deg]                &    119.87 $\pm$     17.31 &    149.61 $\pm$      0.12 \\
        $M_{\rm 0}$ [deg]$^{\rm a}$   &    304.56 $\pm$     17.37 &    255.23 $\pm$      1.05 \\
        \hline
        \noalign{\vskip 0.7mm}
        $a$ [au]                      &       1.09 
        &      13.23 
        \\
        $m \sin i$ [${\rm M}_{\rm jup}$]$^{\rm b}$ &       6.64 
        &     554.94 
        \\
        \hline
        \noalign{\vskip 0.7mm}
        $\sigma_{\rm jit}$ [m\,s$^{-1}$] &    \multicolumn{2}{c}{17.26} \\
        $rms$ [m\,s$^{-1}$]           &     \multicolumn{2}{c}{18.05} \\
        N$_{\rm RV}$ data             &        \multicolumn{2}{c}{191} \\
    \hline \noalign{\vskip 0.7mm}
    \end{tabular}


    \tablefoot{\small $^\mathrm{a}$ The mean anomalies are calculated at the first observational epoch, $t_0 = 2\,451\,482.03 \,\mathrm{JD}$. $^\mathrm{b}$ For the mass and semi-major axis calculation, a primary mass of $m_\mathrm{A} = \unit[1.9 \pm 0.2]{M_\odot}$ is used \citep[see][]{Ortiz2016}.}

    \end{table}

In both reductions, the high-amplitude, long-period variation caused by the stellar companion as well as the low-amplitude, short-period signal induced by the planet are visible. To allow comparison, we modeled each time series with a double-Keplerian model, similarly as described in \citet{Ortiz2016}, using the capabilities of the modeling software for RV and photometry data \texttt{Exostriker} \citep{Trifonov2019}. In each model, a stellar jitter estimate was added quadratically to the individual measurement errors; it was chosen such that the $\chi_\mathrm{red}^2$ of the model becomes 1, leading to jitter values of $\unit[19.8]{m \,s^{-1}}$ for the \texttt{Butler} time series, and $\unit[19.6]{m \,s^{-1}}$ for the \texttt{pyodine} RVs. The best-fit models for both time series are plotted as dashed and dashed-dotted lines in the top plot of Fig.~\ref{fig:LickHIP36616RVs}, and the respective best-fit orbital periods of the two companions are printed in the legend (with uncertainties taken from the covariance matrices of the fits). The periods for each companion agree within their $2 \sigma$-errors, which indicates a high similarity between the two time series.

The bottom two plots of Fig.~\ref{fig:LickHIP36616RVs} display the residuals between the RVs of the two time series and their respective best-fit models. The standard deviation of the \texttt{Butler} residuals is $\unit[19.5]{m \,s^{-1}}$, as compared to $\unit[20.1]{m \,s^{-1}}$ for the \texttt{pyodine} RVs. Two distinct differences in the residuals are visible: First, in the \texttt{Butler} time series a number of RV data points around JD~$2\,454\,500$ follow a systematic downward trend from roughly $\unit[80]{m \,s^{-1}}$ to $\unit[-55]{m \,s^{-1}}$ around the best-fit model, while in \texttt{pyodine} the residuals at that time are much smaller. This trend in the \texttt{Butler} time series is addressed explicitly by \citet{Trifonov2018}, where these exact RVs constrain a dynamical model of mutually inclined orbits between the stellar and planetary companion. However, the authors argue that these data points could as well be noise-induced outliers.

\begin{figure}
\centering
\includegraphics[width=\hsize]{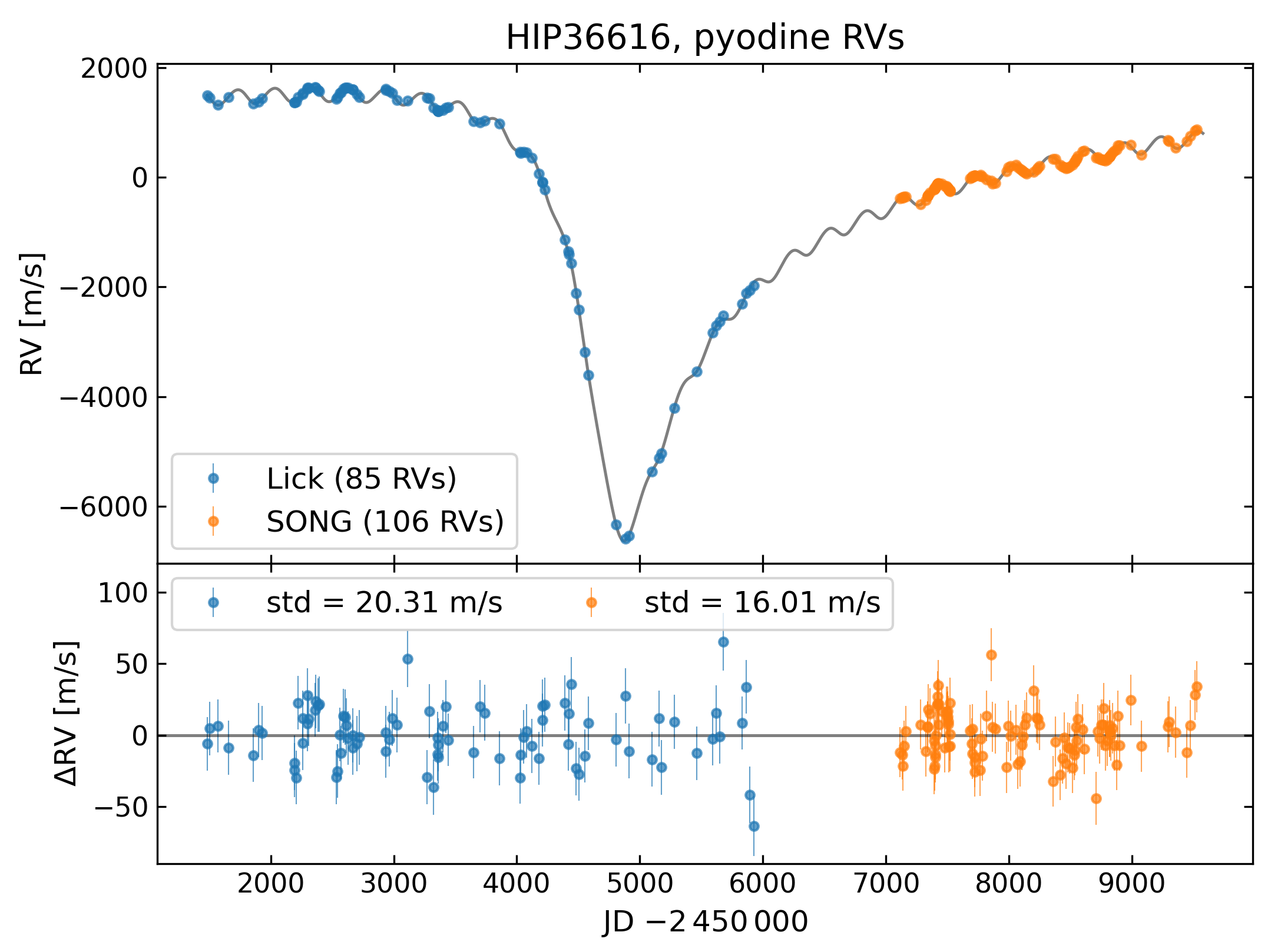}
\caption{RV results of the star HIP~36616, computed with \texttt{pyodine}. \textit{Top:} RVs from Lick (blue) and SONG (orange) spectra, along with the best double-Keplerian fit to the combined RV data sets. \textit{Bottom:} Residuals between the RVs and the best fit.}
\label{fig:HIP36616RVs}
\end{figure}

The second significant difference arises at the very end of the time series, where a few \texttt{pyodine} RVs deviate much more from their best-fit model as compared to the \texttt{Butler} RVs. This could be caused by inaccurate wavelength solutions as discussed above, but it might as well be noise, especially as the last \texttt{pyodine} data point, which shows the largest difference, is not even present in the \texttt{Butler} time series. Still, these differences between the two data sets illustrate the merit of using a second reduction software on the data, as it allows to assess the plausibility of individual RVs.

Furthermore, from March 2015 onward we continued to observe HIP~36616 with the SONG telescope, resulting in another 108 spectra. This allows to take advantage of the flexibility offered by \texttt{pyodine} and compute RVs for the SONG observations as well, which enables us to check the consistency with the Keplerian fit to the Lick RVs. Fig.~\ref{fig:HIP36616RVs}, top, displays the time series determined through \texttt{pyodine} from both Lick and SONG spectra, along with the best double-Keplerian model to the combined RVs. With a jitter value of $\unit[17.3]{m \,s^{-1}}$ applied to both data sets the $\chi_\mathrm{red}^2$ of the fit became 1. The residuals between the RVs and the model are $\unit[20.3]{m\,s^{-1}}$ and $\unit[16.1]{m\,s^{-1}}$ for the Lick and SONG data, respectively, indicating a higher instrumental precision of SONG as compared to Lick.

The best-fit parameters are printed in Table~\ref{table:HIP36616}, and nearly all results fall within the $3\sigma$-uncertainties of the original model from \citet{Ortiz2016}. These results further strengthen the planet hypothesis as a cause of the observed RV variations, and it would be interesting to repeat the dynamical analysis performed by \citet{Trifonov2018} on the complete \texttt{pyodine} time series of Lick and SONG data.

\section{Conclusions and outlook}
\label{sec:Conclusions}

In this work we presented the mathematical algorithms, structure and first results of \texttt{pyodine}, a Python-based code package for the computation of precise RVs from extracted spectra using the I$_2$ cell method. We demonstrated the flexibility of the software by applying it to data from two different instruments, namely the SONG project and the Hamilton spectrograph at Lick observatory. In both cases, \texttt{pyodine} reaches an overall RV precision that matches the ones by the dedicated reduction packages (\texttt{iSONG} and the \texttt{Butler} code, respectively), and we presented RV results of several interesting SONG and Lick targets. 

Particularly on SONG observations of the Sun, \texttt{pyodine} proves its extremely good performance on high-$S/N$ spectra and reaches an estimated short-term precision of $\unit[0.69]{m \, s^{-1}}$. On a long-term time series of the star $\sigma$~Dra however, a strong modulation with roughly yearly period is visible; the modulation is also present (and even more pronounced) in the time series computed with the dedicated instrument software \texttt{iSONG}, and is a known problem of SONG for some stars. It is most probably caused by an instrumental effect and has even been observed in a reduction of the $\sigma$~Dra spectra with the \texttt{Butler} code, so we are safe to assume that it does not point at any problems of \texttt{pyodine} itself.

For a few Lick spectra of the star HIP~36616, we find significant differences between the \texttt{pyodine} and \texttt{Butler} RVs, with each code performing apparently better on some of these spectra, and worse on the others; whether this is caused by noise in the observations that is handled differently by the two reductions, or whether it points at an astrophysical phenomenon, cannot be answered at this point. However, from the whole RV time series of HIP~36616 and results from other Lick targets not presented in this work we can conclude that \texttt{pyodine} roughly matches the RV precision of the \texttt{Butler} code.

Additionally, we presented model results for the Solar-SONG observations of a single chunk centered around three Fe~I-lines. The velocity time series of that single chunk, while being quite noisy, shows clear power excess close to the solar p-mode oscillation frequencies, and this is also true for most other single-chunk velocity time series that we tested. Inspecting single-line velocities such as demonstrated here -- but on larger data sets covering longer baselines to reduce the noise -- might offer additional scientific value: In helio-/asteroseismic analyses this could allow to probe different depths within the atmospheres of the observed targets, possibly aiding a better understanding of their compositions; similarly, by focusing on lines that are strongly affected by stellar activity, this method could help to disentangle Keplerian RV signals from stellar variability. Furthermore, centering chunks specifically around absorption lines can help optimize the efficiency of the code both with respect to RV content and computing power, as parts of the spectrum with low Doppler-content can be neglected and the total number of chunks thus brought down. Particularly for metal-poor stars, which have only few absorption lines and where many of the otherwise automatically generated chunks are poorly constrained, this method might also benefit the final RV precision.

Due to its demonstrated performance and usability and implementation as a Python package, \texttt{pyodine} will be adopted as the standard reduction software for SONG in the future. Furthermore, we plan to implement several upgrades, additions and improvements to the software:

\begin{itemize}
    \item Already in the current version, we have included an experimental routine to compute I$_2$-free spectra from the stellar observations, following the basic recipe described in \citet{Diaz2019}. This can be useful for instance to compute activity indices such as bi-sector spans, and we plan to improve the usability and functionality of the existing routine.
    
    \item Investigating the wavelength-dependence of RVs can help distinguish signals caused by stellar activity from those induced by companions \citep[often called Chromatic Index or CRX, compare][]{Zechmeister2018}. In \texttt{pyodine}, we have incorporated a similar routine which fits the slope of chunk velocities against their central wavelengths to return the CRX of an observation. In the future, we plan to further optimize and test this method to assess its feasibility in the realm of I$_2$ cell reduction codes.
    
    \item In addition to the SONG and Lick instruments presented in this work, \texttt{pyodine} is already being used on spectra from the Waltz telescope at the LSW Heidelberg \citep{Tala2016}. Furthermore, first tests on data from a new SONG node in Mt.\ Kent/Australia delivered promising results, which will be published separately in the near future. It should be straightforward to use \texttt{pyodine} also with other instruments by implementing suitable utility modules; an explanation of the necessary steps to do so is part of the repository's documentation (see information at the end of this section).
    
    \item Finally, we aim to include additional options of sub-models (e.g., other LSF descriptions), to give the user more control and a higher degree of freedom in the analysis.
\end{itemize}

The full \texttt{pyodine} package along with documentation and a minimum working example based on SONG observations is available for download.\footnote{\url{www.gitlab.com/Heeren/pyodine.git}} We invite the scientific community to take part in the future development, for instance through addition of new instruments or suggestions for improvements.

\begin{acknowledgements}
We wish to thank Paul Butler, Debra Fischer, Geoff Marcy and Sharon Wang for many useful conversations and inputs on radial velocity extraction from iodine based data. This work includes observations made with the Hertzsprung SONG telescope operated at the Spanish Observatorio del Teide on the island of Tenerife by the Aarhus and Copenhagen Universities and by the Instituto de Astrofísica de Canarias. 
F.G.\ gratefully acknowledges the work by Mads Fredslund Andersen and the staff at the Observatorio del Teide, leading to the very efficient operation of the Hertzsprung SONG telescope. Funding for the Stellar Astrophysics Center is provided by The Danish National Research Foundation (Grant agreement no. DNRF106). Funding for the Solar-SONG initiative was provided by the Excellence 'Severo Ochoa' Programme at the IAC and the Ministry MINECO under the program PID2019-107187GB-I00. 
Part of this work was supported by the International Max Planck Research School for Astronomy and Cosmic Physics at the University of Heidelberg, IMPRS-HD, Germany. 
S.R.\ and A.Q.\ gratefully acknowledge support of the DFG priority
program SPP 1992 “Exploring the Diversity of Extrasolar Planets” (RE 2694/7-1). 
This research uses services or data provided by the Astro Data Lab at NSF’s NOIRLab. NOIRLab is operated by the Association of Universities for Research in Astronomy (AURA), Inc.\ under a cooperative agreement with the National Science Foundation.
The research has made use of the NASA Exoplanet Archive, which is operated by the California Institute of Technology, under contract with the National Aeronautics and Space Administration under the Exoplanet Exploration Program.
Finally, we thank the anonymous referee for his valuable review.
\end{acknowledgements}

\bibpunct{(}{)}{;}{a}{}{,} 
\bibliographystyle{aa} 
\bibliography{references}

\begin{thebibliography}{50}
\expandafter\ifx\csname natexlab\endcsname\relax\def\natexlab#1{#1}\fi

\bibitem[{{Agard} {et~al.}(1989){Agard}, {Hiraoka}, {Shaw}, \&
  {Sedat}}]{Agard1989}
{Agard}, D., {Hiraoka}, Y., {Shaw}, P., \& {Sedat}, J. 1989, Methods Cell
  Biol., 30, 77

\bibitem[{{Andersen} {et~al.}(2016){Andersen}, {Grundahl}, {Beck}, \&
  {Pall{\'e}}}]{Andersen2016}
{Andersen}, M.~F., {Grundahl}, F., {Beck}, A.~H., \& {Pall{\'e}}, P. 2016, in
  Revista Mexicana de Astronomia y Astrofisica Conference Series, Vol.~48,
  Revista Mexicana de Astronomia y Astrofisica Conference Series, 54--58

\bibitem[{{Andersen} {et~al.}(2014){Andersen}, {Grundahl},
  {Christensen-Dalsgaard}, {Frandsen}, {J{\o}rgensen}, {Kjeldsen}, {Pall{\'e}},
  {Skottfelt}, {S{\o}rensen}, \& {Weiss}}]{Andersen2014}
{Andersen}, M.~F., {Grundahl}, F., {Christensen-Dalsgaard}, J., {et~al.} 2014,
  in Revista Mexicana de Astronomia y Astrofisica Conference Series, Vol.~45,
  83

\bibitem[{{Anglada-Escud{\'e}} \& {Butler}(2012)}]{AngladaEscude2012}
{Anglada-Escud{\'e}}, G. \& {Butler}, R.~P. 2012, \apjs, 200, 15

\bibitem[{Antoci {et~al.}(2013)Antoci, Handler, Grundahl, Carrier, Brugamyer,
  Robertson, Kjeldsen, Kok, Ireland, \& Matthews}]{Antoci2013}
Antoci, V., Handler, G., Grundahl, F., {et~al.} 2013, \mnras, 435, 1563

\bibitem[{{Baranne} {et~al.}(1996){Baranne}, {Queloz}, {Mayor}, {Adrianzyk},
  {Knispel}, {Kohler}, {Lacroix}, {Meunier}, {Rimbaud}, \& {Vin}}]{Baranne1996}
{Baranne}, A., {Queloz}, D., {Mayor}, M., {et~al.} 1996, A\&AS, 119, 373

\bibitem[{{Bedell} {et~al.}(2019){Bedell}, {Hogg}, {Foreman-Mackey}, {Montet},
  \& {Luger}}]{Bedell2019}
{Bedell}, M., {Hogg}, D.~W., {Foreman-Mackey}, D., {Montet}, B.~T., \& {Luger},
  R. 2019, \aj, 158, 164

\bibitem[{{Brahm} {et~al.}(2017){Brahm}, {Jord{\'a}n}, \&
  {Espinoza}}]{Brahm2017}
{Brahm}, R., {Jord{\'a}n}, A., \& {Espinoza}, N. 2017, \pasp, 129, 034002

\bibitem[{{Breton} {et~al.}(2022){Breton}, {Pall{\'e}}, {Garc{\'\i}a},
  {Fredslund Andersen}, {Grundahl}, {Christensen-Dalsgaard}, {Kjeldsen}, \&
  {Mathur}}]{Breton2022}
{Breton}, S.~N., {Pall{\'e}}, P.~L., {Garc{\'\i}a}, R.~A., {et~al.} 2022, \aap,
  658, A27

\bibitem[{Butler {et~al.}(1996)Butler, Marcy, Williams, McCarthy, Dosanjh, \&
  Vogt}]{Butler1996}
Butler, R.~P., Marcy, G.~W., Williams, E., {et~al.} 1996, \pasp, 108, 500

\bibitem[{{Butler} {et~al.}(2017){Butler}, {Vogt}, {Laughlin}, {Burt},
  {Rivera}, {Tuomi}, {Teske}, {Arriagada}, {Diaz}, {Holden}, \&
  {Keiser}}]{Butler2017}
{Butler}, R.~P., {Vogt}, S.~S., {Laughlin}, G., {et~al.} 2017, \aj, 153, 208

\bibitem[{{Campbell} \& {Walker}(1979)}]{Campbell1979}
{Campbell}, B. \& {Walker}, G.~A.~H. 1979, \pasp, 91, 540

\bibitem[{{Cleveland}(1979)}]{Cleveland1979}
{Cleveland}, W.~S. 1979, Journal of the American Statistical Association, 74,
  829

\bibitem[{{Corsaro} {et~al.}(2012){Corsaro}, {Grundahl}, {Leccia}, {Bonanno},
  {Kjeldsen}, \& {Patern{\`o}}}]{Corsaro2012}
{Corsaro}, E., {Grundahl}, F., {Leccia}, S., {et~al.} 2012, \aap, 537, A9

\bibitem[{{Crilly} {et~al.}(2002){Crilly}, {Bernardi}, {Jansson}, \& {da
  Silva}}]{Crilly2002}
{Crilly}, P.~B., {Bernardi}, A., {Jansson}, P.~A., \& {da Silva}, L. E.~B.
  2002, IEEE Transactions on Instrumentation and Measurement, 51, 1142

\bibitem[{{D{\'\i}az} {et~al.}(2019){D{\'\i}az}, {Shectman}, {Butler}, \&
  {Jenkins}}]{Diaz2019}
{D{\'\i}az}, M.~R., {Shectman}, S.~A., {Butler}, R.~P., \& {Jenkins}, J.~S.
  2019, \aj, 157, 204

\bibitem[{{Fredslund Andersen} {et~al.}(2019{\natexlab{a}}){Fredslund
  Andersen}, {Handberg}, {Weiss}, {Frand sen}, {Sim{\'o}n-D{\'\i}az},
  {Grundahl}, \& {Pall{\'e}}}]{Andersen2019}
{Fredslund Andersen}, M., {Handberg}, R., {Weiss}, E., {et~al.}
  2019{\natexlab{a}}, \pasp, 131, 045003

\bibitem[{{Fredslund Andersen} {et~al.}(2019{\natexlab{b}}){Fredslund
  Andersen}, {Pall{\'e}}, {Jessen-Hansen}, {Wang}, {Grundahl}, {Bedding}, {Roca
  Cortes}, {Yu}, {Mathur}, {Gacia}, {Arentoft}, {R{\'e}gulo}, {Tronsgaard},
  {Kjeldsen}, \& {Christensen-Dalsgaard}}]{Andersen2019sun}
{Fredslund Andersen}, M., {Pall{\'e}}, P., {Jessen-Hansen}, J., {et~al.}
  2019{\natexlab{b}}, \aap, 623, L9

\bibitem[{Frink {et~al.}(2002)Frink, Mitchell, Quirrenbach, Fischer, Marcy, \&
  Butler}]{Frink2002}
Frink, S., Mitchell, D.~S., Quirrenbach, A., {et~al.} 2002, \apj, 576, 478

\bibitem[{Frink {et~al.}(2001)Frink, Quirrenbach, Fischer, Röser, \&
  Schilbach}]{Frink2001}
Frink, S., Quirrenbach, A., Fischer, D., Röser, S., \& Schilbach, E. 2001,
  \pasp, 113, 173

\bibitem[{{Gilliland} {et~al.}(1992){Gilliland}, {Morris}, {Weymann}, {Ebbets},
  \& {Lindler}}]{Gilliland1992}
{Gilliland}, R.~L., {Morris}, S.~L., {Weymann}, R.~J., {Ebbets}, D.~C., \&
  {Lindler}, D.~J. 1992, \pasp, 104, 367

\bibitem[{{Griffin}(1973)}]{Griffin1973}
{Griffin}, R. 1973, \mnras, 162, 243

\bibitem[{{Grundahl} {et~al.}(2017){Grundahl}, {Fredslund Andersen},
  {Christensen-Dalsgaard}, {Antoci}, {Kjeldsen}, {Handberg}, {Houdek},
  {Bedding}, {Pall{\'e}}, {Jessen-Hansen}, {Silva Aguirre}, {White},
  {Frandsen}, {Albrecht}, {Andersen}, {Arentoft}, {Brogaard}, {Chaplin},
  {Harps{\o}e}, {J{\o}rgensen}, {Karovicova}, {Karoff}, {Kj{\ae}rgaard
  Rasmussen}, {Lund}, {Sloth Lundkvist}, {Skottfelt}, {Norup S{\o}rensen},
  {Tronsgaard}, \& {Weiss}}]{Grundahl2017}
{Grundahl}, F., {Fredslund Andersen}, M., {Christensen-Dalsgaard}, J., {et~al.}
  2017, \apj, 836, 142

\bibitem[{{Grundahl} {et~al.}(2007){Grundahl}, {Kjeldsen},
  {Christensen-Dalsgaard}, {Arentoft}, \& {Frandsen}}]{Grundahl2007}
{Grundahl}, F., {Kjeldsen}, H., {Christensen-Dalsgaard}, J., {Arentoft}, T., \&
  {Frandsen}, S. 2007, Communications in Asteroseismology, 150, 300

\bibitem[{{Hinkle} {et~al.}(2000){Hinkle}, {Wallace}, {Valenti}, \&
  {Harmer}}]{Hinkle2000}
{Hinkle}, K., {Wallace}, L., {Valenti}, J., \& {Harmer}, D. 2000, {Visible and
  Near Infrared Atlas of the Arcturus Spectrum 3727-9300 {\r{A}}}

\bibitem[{Kanodia \& Wright(2018)}]{Kanodia2018}
Kanodia, S. \& Wright, J. 2018, Research Notes of the {AAS}, 2, 4

\bibitem[{{Keenan} \& {McNeil}(1989)}]{Keenan1989}
{Keenan}, P.~C. \& {McNeil}, R.~C. 1989, \apjs, 71, 245

\bibitem[{{Marcy} \& {Butler}(1992)}]{MarcyButler1992}
{Marcy}, G.~W. \& {Butler}, R.~P. 1992, \pasp, 104, 270

\bibitem[{{Mayor} {et~al.}(2003){Mayor}, {Pepe}, {Queloz}, {Bouchy},
  {Rupprecht}, {Lo Curto}, {Avila}, {Benz}, {Bertaux}, {Bonfils}, {Dall},
  {Dekker}, {Delabre}, {Eckert}, {Fleury}, {Gilliotte}, {Gojak}, {Guzman},
  {Kohler}, {Lizon}, {Longinotti}, {Lovis}, {Megevand}, {Pasquini}, {Reyes},
  {Sivan}, {Sosnowska}, {Soto}, {Udry}, {van Kesteren}, {Weber}, \&
  {Weilenmann}}]{Mayor2003}
{Mayor}, M., {Pepe}, F., {Queloz}, D., {et~al.} 2003, The Messenger, 114, 20

\bibitem[{{McKerns} {et~al.}(2011){McKerns}, {Strand}, {Sullivan}, {Fang}, \&
  {Aivazis}}]{McKerns2011}
{McKerns}, M.~M., {Strand}, L., {Sullivan}, T., {Fang}, A., \& {Aivazis},
  M.~A.~G. 2011, Proceedings of the 10th Python in Science Conference
  [\eprint[arXiv]{1202.1056}]

\bibitem[{{Oja}(1984)}]{Oja1984}
{Oja}, T. 1984, \aaps, 57, 357

\bibitem[{{Ortiz} {et~al.}(2016){Ortiz}, {Reffert}, {Trifonov}, {Quirrenbach},
  {Mitchell}, {Nowak}, {Buenzli}, {Zimmerman}, {Bonnefoy}, {Skemer},
  {Defr{\`e}re}, {Lee}, {Fischer}, \& {Hinz}}]{Ortiz2016}
{Ortiz}, M., {Reffert}, S., {Trifonov}, T., {et~al.} 2016, \aap, 595, A55

\bibitem[{{Pall{\'e}} {et~al.}(2013){Pall{\'e}}, {Grundahl}, {Trivi{\~n}o
  Hage}, {Christensen-Dalsgaard}, {Frandsen}, {Garc{\'\i}a}, {Uytterhoeven},
  {Andersen}, {Rasmussen}, {S{\o}rensen}, {Kjeldsen}, {Spano}, {Nilsson},
  {Hartman}, {J{\o}rgensen}, {Skottfelt}, {Harps{\o}e}, \&
  {Andersen}}]{Palle2013}
{Pall{\'e}}, P.~L., {Grundahl}, F., {Trivi{\~n}o Hage}, A., {et~al.} 2013, in
  Journal of Physics Conference Series, Vol. 440, Journal of Physics Conference
  Series, 012051

\bibitem[{{Pepe} {et~al.}(2021){Pepe}, {Cristiani}, {Rebolo}, {Santos},
  {Dekker}, {Cabral}, {Di Marcantonio}, {Figueira}, {Lo Curto}, {Lovis},
  {Mayor}, {M{\'e}gevand}, {Molaro}, {Riva}, {Zapatero Osorio}, {Amate},
  {Manescau}, {Pasquini}, {Zerbi}, {Adibekyan}, {Abreu}, {Affolter}, {Alibert},
  {Aliverti}, {Allart}, {Allende Prieto}, {{\'A}lvarez}, {Alves}, {Avila},
  {Baldini}, {Bandy}, {Barros}, {Benz}, {Bianco}, {Borsa}, {Bourrier},
  {Bouchy}, {Broeg}, {Calderone}, {Cirami}, {Coelho}, {Conconi}, {Coretti},
  {Cumani}, {Cupani}, {D'Odorico}, {Damasso}, {Deiries}, {Delabre},
  {Demangeon}, {Dumusque}, {Ehrenreich}, {Faria}, {Fragoso}, {Genolet},
  {Genoni}, {G{\'e}nova Santos}, {Gonz{\'a}lez Hern{\'a}ndez}, {Hughes},
  {Iwert}, {Kerber}, {Knudstrup}, {Landoni}, {Lavie}, {Lillo-Box}, {Lizon},
  {Maire}, {Martins}, {Mehner}, {Micela}, {Modigliani}, {Monteiro}, {Monteiro},
  {Moschetti}, {Murphy}, {Nunes}, {Oggioni}, {Oliveira}, {Oshagh}, {Pall{\'e}},
  {Pariani}, {Poretti}, {Rasilla}, {Rebord{\~a}o}, {Redaelli}, {Santana
  Tschudi}, {Santin}, {Santos}, {S{\'e}gransan}, {Schmidt}, {Segovia},
  {Sosnowska}, {Sozzetti}, {Sousa}, {Span{\`o}}, {Su{\'a}rez Mascare{\~n}o},
  {Tabernero}, {Tenegi}, {Udry}, \& {Zanutta}}]{Pepe2021}
{Pepe}, F., {Cristiani}, S., {Rebolo}, R., {et~al.} 2021, \aap, 645, A96

\bibitem[{{Pepe} {et~al.}(2002){Pepe}, {Mayor}, {Galland}, {Naef}, {Queloz},
  {Santos}, {Udry}, \& {Burnet}}]{Pepe2002}
{Pepe}, F., {Mayor}, M., {Galland}, F., {et~al.} 2002, \aap, 388, 632

\bibitem[{{Piskunov} \& {Valenti}(2002)}]{Piskunov2002}
{Piskunov}, N.~E. \& {Valenti}, J.~A. 2002, \aap, 385, 1095

\bibitem[{{Quirrenbach} {et~al.}(2016){Quirrenbach}, {Amado}, {Caballero},
  {Mundt}, {Reiners}, {Ribas}, {Seifert}, {Abril}, {Aceituno},
  {Alonso-Floriano}, {Anwand-Heerwart}, {Azzaro}, {Bauer}, {Barrado},
  {Becerril}, {Bejar}, {Benitez}, {Berdinas}, {Brinkm{\"o}ller}, {Cardenas},
  {Casal}, {Claret}, {Colom{\'e}}, {Cortes-Contreras}, {Czesla}, {Doellinger},
  {Dreizler}, {Feiz}, {Fernandez}, {Ferro}, {Fuhrmeister}, {Galadi},
  {Gallardo}, {G{\'a}lvez-Ortiz}, {Garcia-Piquer}, {Garrido}, {Gesa},
  {G{\'o}mez Galera}, {Gonz{\'a}lez Hern{\'a}ndez}, {Gonzalez Peinado},
  {Gr{\"o}zinger}, {Gu{\`a}rdia}, {Guenther}, {de Guindos}, {Hagen}, {Hatzes},
  {Hauschildt}, {Helmling}, {Henning}, {Hermann}, {Hern{\'a}ndez Arabi},
  {Hern{\'a}ndez Casta{\~n}o}, {Hern{\'a}ndez Hernando}, {Herrero}, {Huber},
  {Huber}, {Huke}, {Jeffers}, {de Juan}, {Kaminski}, {Kehr}, {Kim}, {Klein},
  {Kl{\"u}ter}, {K{\"u}rster}, {Lafarga}, {Lara}, {Lamert}, {Laun},
  {Launhardt}, {Lemke}, {Lenzen}, {Llamas}, {Lopez del Fresno},
  {L{\'o}pez-Puertas}, {L{\'o}pez-Santiago}, {Lopez Salas}, {Magan
  Madinabeitia}, {Mall}, {Mandel}, {Mancini}, {Marin Molina}, {Maroto
  Fern{\'a}ndez}, {Mart{\'\i}n}, {Mart{\'\i}n-Ruiz}, {Marvin}, {Mathar},
  {Mirabet}, {Montes}, {Morales}, {Morales Mu{\~n}oz}, {Nagel}, {Naranjo},
  {Nowak}, {Palle}, {Panduro}, {Passegger}, {Pavlov}, {Pedraz}, {Perez},
  {P{\'e}rez-Medialdea}, {Perger}, {Pluto}, {Ram{\'o}n}, {Rebolo}, {Redondo},
  {Reffert}, {Reinhart}, {Rhode}, {Rix}, {Rodler}, {Rodr{\'\i}guez},
  {Rodr{\'\i}guez L{\'o}pez}, {Rohloff}, {Rosich}, {Sanchez Carrasco},
  {Sanz-Forcada}, {Sarkis}, {Sarmiento}, {Sch{\"a}fer}, {Schiller}, {Schmidt},
  {Schmitt}, {Sch{\"o}fer}, {Schweitzer}, {Shulyak}, {Solano}, {Stahl},
  {Storz}, {Tabernero}, {Tala}, {Tal-Or}, {Ulbrich}, {Veredas}, {Vico Linares},
  {Vilardell}, {Wagner}, {Winkler}, {Zapatero Osorio}, {Zechmeister},
  {Ammler-von Eiff}, {Anglada-Escud{\'e}}, {del Burgo}, {Garcia-Vargas},
  {Klutsch}, {Lizon}, {Lopez-Morales}, {Ofir}, {P{\'e}rez-Calpena}, {Perryman},
  {S{\'a}nchez-Blanco}, {Strachan}, {St{\"u}rmer}, {Su{\'a}rez}, {Trifonov},
  {Tulloch}, \& {Xu}}]{Quirrenbach2016}
{Quirrenbach}, A., {Amado}, P.~J., {Caballero}, J.~A., {et~al.} 2016, in
  Society of Photo-Optical Instrumentation Engineers (SPIE) Conference Series,
  Vol. 9908, Ground-based and Airborne Instrumentation for Astronomy VI, ed.
  C.~J. {Evans}, L.~{Simard}, \& H.~{Takami}, 990812

\bibitem[{Reffert {et~al.}(2015)Reffert, Bergmann, Quirrenbach, Trifonov, \&
  Künstler}]{Reffert2015}
Reffert, S., Bergmann, C., Quirrenbach, A., Trifonov, T., \& Künstler, A.
  2015, \aap, 574, A116

\bibitem[{Stetson(1989)}]{stetson1989techniques}
Stetson, P. 1989, The Techniques of Least Squares and Stellar Photometry with
  CCDs: A Series of Five Lectures Presented At: V Escola Avan{\c{c}}ada de
  Astrof{\'\i}sica, 1989 July 30 - August 3, Aguas de S{\~a}o Pedro, Brazil

\bibitem[{{Tala} {et~al.}(2016){Tala}, {Heeren}, {Grill}, {Harris},
  {St{\"u}rmer}, {Schwab}, {Gutcke}, {Reffert}, {Quirrenbach}, {Seifert},
  {Mandel}, {Geuer}, {Sch{\"a}ffner}, {Thimm}, {Seeman}, {Tietz}, \&
  {Wagner}}]{Tala2016}
{Tala}, M., {Heeren}, P., {Grill}, M., {et~al.} 2016, in Society of
  Photo-Optical Instrumentation Engineers (SPIE) Conference Series, Vol. 9908,
  Ground-based and Airborne Instrumentation for Astronomy VI, ed. C.~J.
  {Evans}, L.~{Simard}, \& H.~{Takami}, 99086O

\bibitem[{{Trifonov}(2019)}]{Trifonov2019}
{Trifonov}, T. 2019, {The Exo-Striker: Transit and radial velocity interactive
  fitting tool for orbital analysis and N-body simulations}

\bibitem[{{Trifonov} {et~al.}(2018){Trifonov}, {Lee}, {Reffert}, \&
  {Quirrenbach}}]{Trifonov2018}
{Trifonov}, T., {Lee}, M.~H., {Reffert}, S., \& {Quirrenbach}, A. 2018, \aj,
  155, 174

\bibitem[{{Valenti} {et~al.}(1995){Valenti}, {Butler}, \&
  {Marcy}}]{Valenti1995}
{Valenti}, J.~A., {Butler}, R.~P., \& {Marcy}, G.~W. 1995, \pasp, 107, 966

\bibitem[{{van Leeuwen}(2007)}]{vanLeeuwen2007}
{van Leeuwen}, F. 2007, \aap, 474, 653

\bibitem[{{Vogt}(1987)}]{Vogt1987}
{Vogt}, S.~S. 1987, \pasp, 99, 1214

\bibitem[{{Vogt} {et~al.}(2014){Vogt}, {Radovan}, {Kibrick}, {Butler},
  {Alcott}, {Allen}, {Arriagada}, {Bolte}, {Burt}, {Cabak}, {Chloros},
  {Cowley}, {Deich}, {Dupraw}, {Earthman}, {Epps}, {Faber}, {Fischer}, {Gates},
  {Hilyard}, {Holden}, {Johnston}, {Keiser}, {Kanto}, {Katsuki}, {Laiterman},
  {Lanclos}, {Laughlin}, {Lewis}, {Lockwood}, {Lynam}, {Marcy}, {McLean},
  {Miller}, {Misch}, {Peck}, {Pfister}, {Phillips}, {Rivera}, {Sandford},
  {Saylor}, {Stover}, {Thompson}, {Walp}, {Ward}, {Wareham}, {Wei}, \&
  {Wright}}]{Vogt2014}
{Vogt}, S.~S., {Radovan}, M., {Kibrick}, R., {et~al.} 2014, \pasp, 126, 359

\bibitem[{{Wang} {et~al.}(2020){Wang}, {Wright}, {MacQueen}, {Cochran}, {Doss},
  {Gibson}, \& {Schmitt}}]{Wang2020}
{Wang}, S.~X., {Wright}, J.~T., {MacQueen}, P., {et~al.} 2020, \pasp, 132,
  014503

\bibitem[{{Wright} \& {Eastman}(2014)}]{Wright2014}
{Wright}, J.~T. \& {Eastman}, J.~D. 2014, \pasp, 126, 838

\bibitem[{{Zechmeister} \& {K{\"u}rster}(2009)}]{Zechmeister2009}
{Zechmeister}, M. \& {K{\"u}rster}, M. 2009, \aap, 496, 577

\bibitem[{{Zechmeister} {et~al.}(2018){Zechmeister}, {Reiners}, {Amado},
  {Azzaro}, {Bauer}, {B{\'e}jar}, {Caballero}, {Guenther}, {Hagen}, {Jeffers},
  {Kaminski}, {K{\"u}rster}, {Launhardt}, {Montes}, {Morales}, {Quirrenbach},
  {Reffert}, {Ribas}, {Seifert}, {Tal-Or}, \& {Wolthoff}}]{Zechmeister2018}
{Zechmeister}, M., {Reiners}, A., {Amado}, P.~J., {et~al.} 2018, \aap, 609, A12

\end{thebibliography}


\begin{appendix} 

\section{\texttt{iSONG} results of SONG Solar data}
\label{app:1}

\begin{figure}[th!]
\centering
\includegraphics[width=\textwidth]{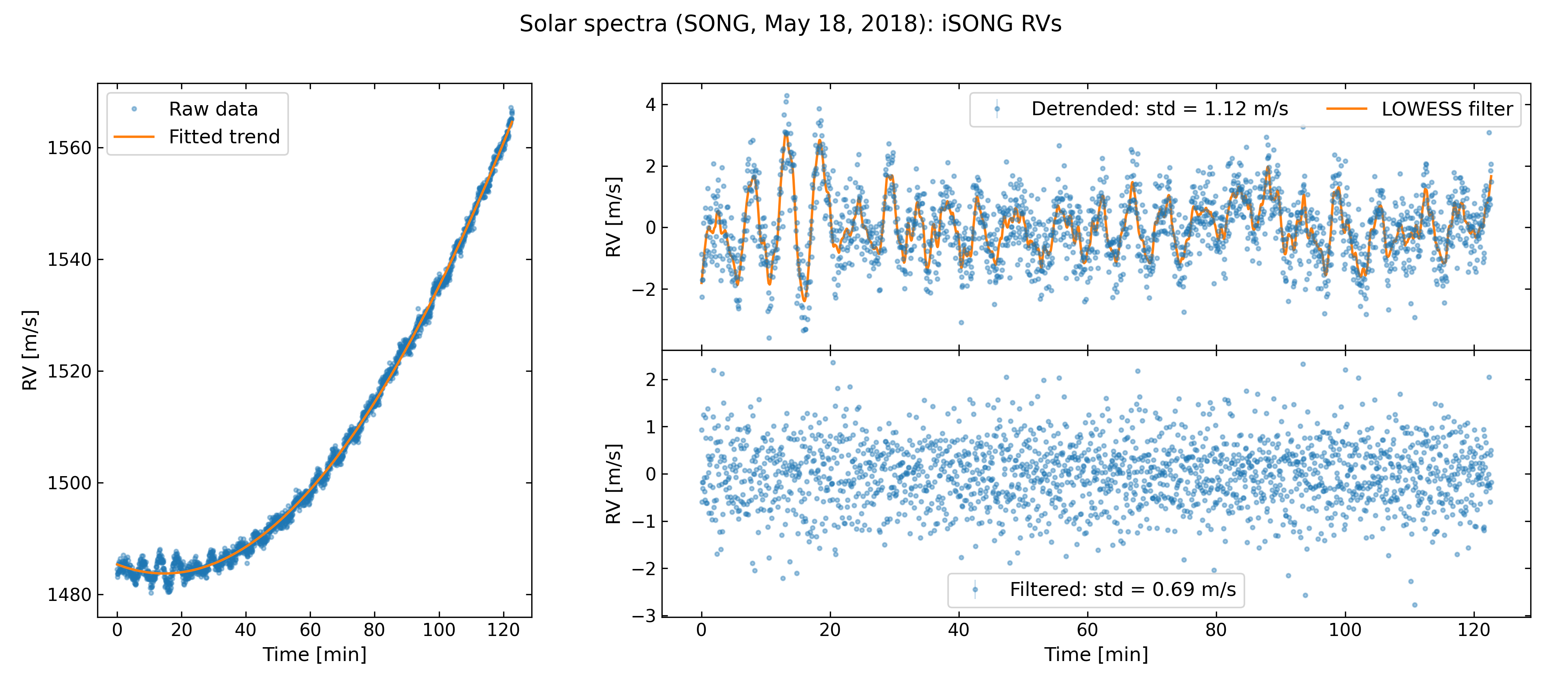}
\caption{Same as Fig.~\ref{fig:SONGsunRVs}, but using the RV results produced by \texttt{iSONG}.}
\label{fig:SONGsunRVs_iSONG}
\end{figure}

\FloatBarrier

\end{appendix}

\end{document}